\documentclass[twocolumn,showpacs,amsfonts,amssymb,amsmath,floats,aps]{revtex4-1}
\usepackage{graphicx}
\usepackage{dcolumn}
\usepackage{multirow}
\def\expect#1{\left\langle#1\right\rangle}

\def\ket#1{\left|#1\right\rangle}
\def\braket#1#2#3{\left\langle#1\middle|#2\middle|#3\right\rangle}
\def\mueff{\mu^2_{\rm eff}}
\def\Qeff{Q^2_{\rm eff}}
\def\ie{i.e.\ }
\def\etal{{\em et al.}~}
\def\e{\varepsilon}
\def\ed{\epsilon_{\rm d}}
\def\edi{\epsilon_{\rm d,i}}
\def\edf{\epsilon_{\rm d,f}}
\def\nd{n_{\rm d}}
\def\Hf{H_{\rm f}}
\def\Hi{H_{\rm i}}
\def\Gn{\Gamma_{\rm 0}}

\def\Geff{\Gamma_{\rm eff}}
\def\gc{g_{\rm c}}

\def\gs{g^*}
\def\gp{g'}
\def\gi{g_{\rm i}}
\def\gf{g_{\rm f}}

\def\Uf{U_{\rm f}}
\def\Uc{U_{\rm c}}
\def\Uren{U_{\rm ren}}
\def\oc{\omega_{\rm c}}

\def\Deq{\expect{D}_{\rm eq}}
\def\Dt{\expect{D(t)}}
\def\nd{n_{\rm d}}
\def\ndeq{\expect{\nd}_{\rm eq}}
\def\ndt{\expect{\nd(t)}}
\def\tco{t_{\rm co}}

\def\Nz{N_{\rm z}}
\def\Nb{N_{\rm B}}
\def\Ns{N_{\rm S}}
\def\w{\omega}
\def\non{\nonumber \\ }

\begin{document}

\title{Influence of a bosonic environment onto the non-equilibrium dynamics of
local electronic states in a quantum impurity system close to a quantum phase
transition}
\author{Christian Kleine}
\author{Frithjof B. Anders}
\address{Lehrstuhl f\"{u}r Theoretische Physik II, Technische Universit\"{a}t
Dortmund, Otto-Hahn-Stra{\ss}e 4, 44221 Dortmund, Germany}
\date{\today}

\begin{abstract}
We investigate the influence of an additional bosonic bath onto the real-time
dynamics of a localized orbital coupled to conduction band with an
energy-dependent coupling function $\Gamma(\e) \propto |\e|^r$.
Recently, a rich phase diagram has been found in this Bose-Fermi Anderson model,
where the transitions between competing ground states are governed by
quantum critical points.
In addition to a transition between a Kondo singlet and a local moment,
a localized phase has been established once the coupling to a sub-ohmic
bosonic bath exceeds a critical value.
Using the time-dependent numerical renormalization group approach,
we show that the non-equilibrium dynamics with F-type of bath exponents can
be fully understand within an effective single-impurity Anderson model using
a renormalized local Coulomb interaction $\Uren$.
For regimes with B-type of bath exponents, the nature of the bosonic bath and
the coupling strength has a profound impact on the electron dynamics which
can only partially be understood using an appropriate $\Uren$.
The local expectation values always reach a steady state at very long times.
By a scaling analysis for $\Lambda \to 1^+$, we find thermalization of the
system only in the strong coupling regime.
In the local moment and in the localized phase significant deviations between
the steady-state value and the thermal equilibrium value are found that are
related to the distance to the quantum critical point.
\end{abstract}
\pacs{05.70.Ln,72.15.Qm,78.67.Hc}
\maketitle

\section{Introduction}

Quantum impurity systems (QISs) \cite{BullaCostiPruschke2008} are of great
interest for the understanding of nanoscale devices, such as semi-conductor
quantum dots, or qubits for quantum computation \cite{NielsenChuang2010}.
These systems consist of a small subsystem with a finite number of
degrees of freedom (DOFs), interacting with an infinitely large environment of
non-interacting particles.
Several classes of models have been investigated with purely fermionic
\cite{Wilson1975,KrishnaWilkinsWilson1980,KrishnaWilkinsWilson1980b,
BullaPruschkeHewson1997,GonzalezIngersent1998}
or purely bosonic
\cite{LeggettChakravartyDorseyFisherGargZwerger1987,BullaTongVojta2003,
BullaLeeHyunNingVojta2005}
environments in the last decades.
In many of these models impurity quantum phase transitions (QPTs)
have been found and are well understood \cite{Vojta2006}.
In contrary to bulk QPTs \cite{Stewart01}, only a subset of the DOFs becomes
critical in these boundary QPTs \cite{Vojta2006}.

In the last decade one focus has been on mixed Bose-Fermi QISs containing
multiple baths of fermionic and bosonic DOFs
\cite{KircanVojta2004,Vojta2006,GlossopIngersent2005,GlossopIngersent2007,
ChungGlossopFritzKircanIngersent2007,GlossopKhoshkhouIngersent2008,
ChengGlossopIngersent2009,NicaIngersentZhuSi2013,PixleyKirchnerIngersentSi2013}
extending some earlier work \cite{Haldane1977,Haldane1977b,Freericks1993}.
The spin-$1/2$ Bose-Fermi Kondo model (BFKM) \cite{GlossopIngersent2007}
and its extension to the charge sector, the Bose-Fermi Anderson model (BFAM)
\cite{ChengGlossopIngersent2009,PixleyKirchnerIngersentSi2013},
are the most studied models.
These models can serve as effective site in the extended dynamical mean-field
theory (EDMFT) \cite{SiRabelloIngersentSmith2001,SiRabelloIngersentSmith2003}
which is used to address the occurrence of magnetic QPTs in heavy fermions
\cite{Grewe91,Stewart01} and in the Kondo lattice model (KLM)
\cite{KLM-Sigrist91,Coleman1999}.
Hereby, the bosonic parts of the BFAM represent the fluctuating effective
magnetic order parameter field generated by the other $f$ moments.
In the KLM the magnetic ordering induced by the Ruderman-Kittel-Kasuya-Yosida
(RKKY) interaction \cite{RudermanKittel1954,Kasuya1956,Yosida1957,Fazekas1999}
competes with the screening of the $f$ shell moments due to the Kondo
effect \cite{Kondo1964,Doniach77}.

The model has been applied to the equilibrium properties of a noisy quantum-dot
system \cite{Hur2004,LiHur2005} where the fluctuations of the gate voltage
provide the bosonic environment.
Investigating the non-equilibrium dynamics of QISs is essential for the
understanding of relaxation and dissipation for these nano devices and qubits.

In the weak coupling limit, rate equations and Born-Markov approaches
have been used \cite{MayKuehn2000,welack-2006-124} to explore the real-time
dynamics and the steady state.
Such methods, however, cannot be employed in the strong coupling limit or
close to a quantum critical point (QCP) where competing orthogonal ground
states need to be accounted for.
In this regime, the non-equilibrium extension of Wilson's numerical
renormalization group (NRG) approach \cite{Wilson75,BullaCostiPruschke2008},
the time-dependent numerical renormalization group (TD-NRG)
\cite{AndersSchiller2005,AndersSchiller2006}, has been used to access the
real-time dynamics in QISs with purely fermionic or bosonic baths
\cite{AndersSchiller2005,AndersSchiller2006,AndersBullaVojta2007,
RoosenHofstetter2008,NghiemCosti2014,LechtenbergAnders2014,
KleineMusshoffAnders2014} or steady-state currents through nano devices
\cite{AndersSSnrg2008,SchmittAnders2011,JovchevAnders2013}.
Recently, real-time quantum Monte-Carlo approaches \cite{Rabani08,
SchmidtWerner2008,GulletAl2011,SchiroFabrizio2009}
and multilayer multiconfiguration time-dependent Hartree methods
\cite{MuehlbacherThoss2012,WilnerThossRabani2013} have also been successfully
applied to such problems as well as the density matrix renormalization group,
see Ref.~\cite{Schollwoeck2011} for a recent review.

This paper focuses on the real-time dynamics of an Anderson impurity coupled
to a bosonic as well as a fermionic bath.
For that purpose, we have extended the recently introduced Bose-Fermi NRG
\cite{GlossopIngersent2007} to non-equilibrium using the TD-NRG.
The BFAM and the sub-ohmic spin-boson model (SBM) belong to the same
universality class \cite{GlossopIngersent2007,ChengGlossopIngersent2009,
PixleyKirchnerIngersentSi2013} for an ungaped fermionic coupling function.
It has been shown that the BFAM exhibits a complicated phase diagram
\cite{GlossopIngersent2007,ChengGlossopIngersent2009,
PixleyKirchnerIngersentSi2013} for a pseudo-gap density of states (DOS)
whose low-energy properties are characterized by the exponent $r$:
$\rho(\e) \propto |\e|^r$.
Although $r = 1$ is relevant for quantum impurities in $d$ wave superconductors
and charge-neutral graphene at low energies, we use $r$ as a model parameter
only in the range $0 \leq r < 1/2$ since in this regime QPTs between a
free local moment phase and a screened magnetic moment phase are found
due to the Kondo effect.
The low-frequency part of the bosonic bath coupling function $J(\w)$
can also be approximated by a power-law \cite{Leggett1987}:
$J(\w) \propto \w^s$.
In this paper we will focus on the sub-ohmic regime $1/2 < s < 1$ to avoid the
NRG deficiency \cite{vojtaErr2009,VojtaBullaGuettgeAnders2010,Vojta2012}
in the regime $0 < s < 1/2$ and we will comment on the case of an
arbitrary $s$.

The equilibrium properties of the BFAM \cite{ChengGlossopIngersent2009}
reveal a QPT with hyperscaling behavior for the above stated regimes for
$r$ and $s$.
The specific combinations of the bath exponents $r$ and $s$ have not only
a profound impact on the equilibrium properties of the QPT, but also divide
different regimes in the real-time dynamics.
We will show that for a final Hamiltonian, whose low-energy fixed point (FP)
is mainly governed by the fermionic bath properties (F-type), the
non-equilibrium dynamics of the BFAM can be exactly reproduced by a purely
fermionic single-impurity Anderson model (SIAM) replacing the bare $U$ by an
$\Uren$ which includes the attractive electron-electron interaction mediated
by the bosonic bath.

For exponents where the low-temperature FP is governed by the bosonic bath
(B-type) \cite{PixleyKirchnerIngersentSi2013}, the real-time dynamics shows
distinctive differences to those in an effective SIAM defined by appropriate
values of $\Uren$ such that the local thermodynamic expectation values are
in both models identical for the initial and final parameters
of the Hamiltonian.
Although, the local thermodynamic expectation values are identical 
the two models can be in two different low-temperature phases:
While the dynamics in the BFAM is driven by the localized phase, the
dynamics in effective SIAM is still governed by the strong coupling FP
leading to a deviating dynamics when comparing the real-time dynamics 
of the same local quantities in both models.

For quenches of the gate voltage, we find different steady-state values for
the impurity charge filling for identical bath couplings depending on
the choice of the initial conditions since the FP of the final Hamiltonian
is in the localized phase.
We will demonstrate thermalization within the numerical accuracy of the
method for Hamiltonians approaching the strong coupling FP.
The microscopic mechanism  of experimentally observed hysteretic behavior
of the I(V) curves \cite{LiHysteresis2003} in molecular junctions when sweeping
the voltage with a very small but finite rate is still unsolved.
It has been suggested that such a behavior might be related to
conformational changes in these complex molecules
\cite{Donhauser2001,GalperinRatnerNitzan2007}.
At infinitely large times and at finite temperatures, a real molecular junction
should approach a unique steady state \cite{Hershfield1993,JovchevAnders2013}.
Using a reduced density matrix formalism for the spinless version of a two-lead
BFAM, Wilner \etal have found a bistability in the real-time evolution of
the molecular electronic occupancy \cite{MuehlbacherThoss2012,
WilnerThossRabani2013} depending on two different initial preparations
of the bosonic bath configuration.
The major difference between a spinless and a spinfull BFAM is the sign of
the effective Coulomb interaction generated by the electron-boson interaction.
In the spinless case, a repulsive interaction between the local electron
density and the local conduction electrons are generated and an effective
interacting resonant level model \cite{VigmanFinkelstein78,Schlottmann1980}
emerges.
The effective equilibrium Coulomb repulsion $\Uren$ can be extracted
\cite{EidelsteinSchiller2013,JovchevAnders2013} and favors coherent
oscillations \cite{GuettgeAndersSchiller2013} in the real-time dynamics.
In contrary, a local attractive Coulomb interaction mediated by bosons
is induced \cite{Mahan81} for models including the spin as investigated
in this paper.
In addition, such models exhibit quantum phase transitions for large
electron-boson couplings
\cite{ChengGlossopIngersent2009,PixleyKirchnerIngersentSi2013}
and might, therefore, also contain bistabilities, depending on the initial
conditions.
In this paper, however, we focus on a spinfull single-lead BFAM.

\subsection{Plan of the paper}

This paper is organized as follows:
In Sec.~\ref{sec:theory} we present the model of interest, the Bose-Fermi
Anderson model (Sec.~\ref{sec:bfam}), and introduce the numerical tools,
the NRG (Sec.~\ref{sec:nrg}) and the TD-NRG (Sec.~\ref{sec:td-nrg}).
To prepare the reader for the dynamics, we summarize the main
equilibrium properties of the system in Sec.~\ref{sec:equilibrium-properties}.
After some general remarks, where we will classify the F-type,
B-type and M-type bath exponent combinations
\cite{ChengGlossopIngersent2009,PixleyKirchnerIngersentSi2013},
we will discuss the generation of an additional attractive Coulomb
interaction due to the coupling to the bosonic bath in
Sec.~\ref{sec:eq-renormalized-coulomb-repulsion}
and provide an overview of the known phases and fixed points in
Sec.~\ref{sec:eq-overview-phases} for the ph-symmetric model.
The equilibrium expectation value of the double occupancy will be
be used to define a mapping between the BFAM and an effective SIAM,
as introduced in Sec.~\ref{sec:eq-double-occupancy}.

In Sec.~\ref{sec:td-nrg-results} we present our results on the
non-equilibrium dynamics in the BFAM.
We distinguish between F-type bath exponents in Sec.~\ref{sec:td-nrg-f-type},
where the dynamics are reproduced exactly by an effective SIAM, and B-type
bath exponents in Sec.~\ref{sec:quenches-bosonic-type}, for which distinctive
differences in the dynamics between the BFAM and SIAM  have been found.
While in those sections ph symmetry remains maintained at any time,
we focus on the influence of a sudden change of a gate voltage,
breaking ph symmetry, on the dynamics of the impurity charge occupancy
in Sec.~\ref{sec:ph-symmetry-broken}.
We end this paper with a brief conclusion.

\section{Theory}
\label{sec:theory}

\subsection{The Bose-Fermi Anderson model}
\label{sec:bfam}

The Bose-Fermi Anderson model consists of an Anderson impurity
\cite{Anderson61} that hybridizes with conduction band electrons and
additionally couples to an ungaped fermionic bath with a continuous
spectrum via its electron density.
The origin of the bosons can be either acoustic phonons \cite{Haldane1977,
Haldane1977b} or bosonized magnetic fluctuations with the extended DMFT
\cite{ChengGlossopIngersent2009,PixleyKirchnerIngersentSi2013}.
In recent years, its equilibrium properties have been extensively studied
using Wilson's NRG \cite{ChengGlossopIngersent2009,
PixleyKirchnerIngersentSi2013}.

The Hamiltonian of the charge-coupled Bose-Fermi Anderson model comprises
several parts:
(i) the free conduction-electron bath
\begin{align}
  H_{\rm F} = \sum_{\vec{k},\sigma} \epsilon_{\vec{k}}
                c_{\vec{k},\sigma}^\dagger
                c_{\vec{k},\sigma}^{\phantom{\dagger}}
\end{align}
where $c_{\vec{k},\sigma}^\dagger$ creates a quasiparticle with the
dispersion $\epsilon_{\vec{k}}$ and spin $\sigma$, and non-interacting
bosons
\begin{align}
  H_{\rm B} = \sum_{\vec{q}} \omega_{\vec{q}} b_{\vec{q}}^\dagger
                                              b_{\vec{q}}^{\phantom{\dagger}}
\end{align}
where $b_{\vec{q}}^\dagger$ creates a bosonic excitation with the
momentum $\vec{q}$ and the dispersion $\omega_{\vec{q}}$, and
(ii) the impurity part of a spin-degenerate impurity level
\begin{align}
  H_{\rm imp} = \ed \left( n_{\rm d \uparrow} + n_{\rm d \downarrow} \right)
                + U n_{\rm d \uparrow} n_{\rm d \downarrow}
  \label{eq:Himp}
\end{align}
in the absence of a magnetic field.
The operator $d_{\sigma}^\dagger$ creates an impurity electron with spin
$\sigma$ and the single-particle energy $\ed$.
The spin-dependent electron density of the impurity is represented by
$n_{\rm d \sigma} = d_{\sigma}^\dagger d_{\sigma}^{\phantom{\dagger}}$, 
and $U$ denotes the local Coulomb repulsion.
(iii) The interaction between the impurity and the conduction band is given by
\begin{align}
  H_{\rm int,F} = \sum_{\vec{k},\sigma} \left(
    V_{\vec{k}} c_{\vec{k} \sigma}^\dagger d_\sigma^{\phantom{\dagger}}
    + V_{\vec{k}} d_{\sigma}^\dagger c_{\vec{k} \sigma}^{\phantom{\dagger}}
  \right)
\end{align}
and the local electron charge couples to each displacement of the bosonic modes
\begin{align}
  H_{\rm int,B} = \left( \sum_\sigma  n_{\rm d \sigma} - 1 \right)
    \sum_{\vec{q}} g_{\vec{q}}
    \left( b_{\vec{q}}^\dagger + b_{\vec{q}}^{\phantom{\dagger}} \right)
\end{align}
via the coupling constants $g_{\vec{q}}$ and its form ensures particle-hole
symmetry.
The full Hamiltonian
\begin{align}
   \label{eq:full-H}
   H_{\rm BFAM} = H_{\rm imp} + H_{\rm F} + H_{\rm B} + H_{\rm int,F}
   + H_{\rm int,B}
\end{align}
determines the thermodynamics and the dynamics of the system.

\subsubsection{Fermionic bath}
\label{sec:baths-fermionic}

In the absence of a bosonic bath the dynamics of the magnetic impurity is
fully determined by the coupling function \cite{KrishnaWilkinsWilson1980,
BullaPruschkeHewson1997,GonzalezIngersent1998,GlossopLogan2003}
\begin{align}
\label{eqn:6}
  \Gamma_\sigma \left( \e \right) = \pi \sum_{\vec{k}} V_{\vec{k}}^2
    \delta \left( \e - \e_{\vec{k}\sigma} \right)
\end{align}
which we will take as spin-independent in the following.
Ignoring the spectral details of its high-energy part, which only influence
the values of the relevant energy scales, but do not determine the low-energy
fixed point spectrum, $\Gamma(\e)$ is replaced by the particle-hole
symmetric power-law \cite{WithoffFradkin1990,BullaPruschkeHewson1997,
GonzalezIngersent1998,GlossopLogan2003} of the form
\begin{align}
  \Gamma \left( \e \right)
    = \Gn \left| \frac{\e}{D} \right|^r
      \Theta \left( D - \left| \e \right| \right)
\end{align}
with the exponent $r$ and the cutoff $D$, which defines the effective bandwidth.
Note that with this definition the integral over the coupling function,
\begin{align}
  \pi V_0^2 = \int \Gamma(\e) \; {\rm d} \e =\frac{2 \Gn D}{r + 1}
  \quad ,
\end{align}
is dependent on the bath exponent $r \ge 0$:
With increasing $r$ the hybridization $V_0$ decreases.

The parameter $\Gn$ serves as energy unit of the problem that turns into
the standard charge fluctuation scale for a constant DOS ($r=0$).
While $r=0$ and $r=1$ are the prototypical experimental realizations
--- $r=1$ is relevant in $d$ wave superconductors or in graphene ---
we take $r$ as an arbitrary parameter of the model.
Furthermore, the particle-hole asymmetry is governed by the parameter
$\Delta \epsilon = 2\epsilon_{\rm d} + U$.

\subsubsection{Bosonic bath}
\label{sec:baths-bosonic}

Decoupling the fermionic bath, the local impurity dynamics
is completely determined by the bosonic bath coupling function
\begin{align}
  \label{eq:J-w-def}
  J \left( \omega \right)
    = \pi \sum_{\vec{q}} g_{\vec{q}}^2
      \delta \left( \omega - \omega_{\vec{q}} \right)
\end{align}
which is approximated by the power-law form
\cite{LeggettChakravartyDorseyFisherGargZwerger1987}
\begin{align}
  \label{eq:J-w-pl}
  J \left( \omega \right)
    = 2 \pi g \omega_c^{1-s} \omega^s \;
      \Theta \left( \omega_c - \omega \right)
\end{align}
for the same reason as in the fermionic case:
The low-frequency properties and the nature of the quantum critical points
are governed by the exponent $s$.
The cutoff $\oc$ defines the high-energy scale while the overall
coupling strength is denoted by the dimensionless coupling $g$.

The exponent $s$ separates different bath types:
$0 < s < 1$ for a sub-ohmic dissipation \cite{BullaLeeHyunNingVojta2005},
$s = 1$ for ohmic dissipation \cite{LeggettChakravartyDorseyFisherGargZwerger1987}
and $s > 1$ for a super-ohmic bath \cite{Weiss1993}.
In the following we restrict ourselves to bosonic exponents
$1/2 \leq s \leq 1$ due to so far unresolved problems applying the NRG
\cite{VojtaBullaGuettgeAnders2010} for the sub-ohmic spin-boson model
with $0 < s < 1/2$.

\subsection{The numerical renormalization group for Bose-Fermi systems}
\label{sec:nrg}

The numerical renormalization group (NRG) is well established for fermionic
systems \cite{Wilson1975,KrishnaWilkinsWilson1980,KrishnaWilkinsWilson1980b}.
In the last decade the NRG has been extended to bosonic environments
\cite{BullaTongVojta2003,BullaLeeHyunNingVojta2005,Bulla2009,
KirchnerIngersentSi2012,Vojta2012} and lately combined for Bose-Fermi
impurity systems \cite{GlossopIngersent2005,GlossopIngersent2007,
ChengGlossopIngersent2009}.
We shortly summarize the NRG procedure \cite{GlossopIngersent2007} 
for an environment consisting of bosonic as well as fermionic DOFs.

The bath continua of the two baths will be \emph{discretized logarithmically}
with the discretization parameter $\Lambda > 1$ \cite{Wilson1975}.
The fermionic conduction band is divided into the intervals
$\lbrack \Lambda^{-(n+1)} D,  \Lambda^{-n} D \rbrack$
and
$\lbrack - \Lambda^{-n} D, - \Lambda^{-(n+1)} D \rbrack$
regarding the bandwidth $D$ and covers positive and negative energies.
The bosonic bath is divided into the intervals
$\lbrack \Lambda^{-(m+1)} \oc, \Lambda^{-m} \oc \rbrack$
regarding the high-energy cutoff $\oc$ and covers only positive energies.
The indexes $n,m=0,1,2,...$ enumerate the fermionic and bosonic intervals.
In each interval only one discrete energy is used as representative,
which is the main approximation of the NRG \cite{Wilson1975,
BullaCostiPruschke2008}.
For simplicity we set both bath cut-offs equal: $D / \oc = 1$.

This assumption introduced by Glossop \etal \cite{GlossopIngersent2007}
is well justified when targeting the low-energy physics and quantum phase
transitions.
One could start with $D_0 \gg D$ and apply an RG procedure to the problem
until the effective bandwidth $D_{\rm eff} \approx \oc$ is reached.
This would give rise to an effective impurity model with renormalized
parameters ${\ed}_{\rm , eff}$ and $U_{\rm eff}$ as well as sightly modified 
$\Geff(\w)$ entering Eq.~\eqref{eqn:6}.

\begin{figure}[b]
  \includegraphics[width=0.9\linewidth]{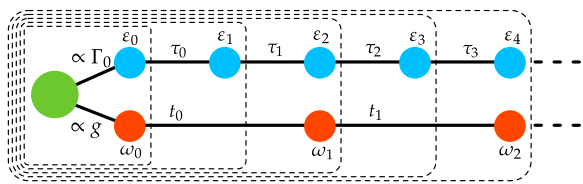}
  \caption{\label{fig:wilsonchain}%
    (Color online)
    Sketch of the discretized model.
    The impurity (green) only couples to the first sites of the two
    independent Wilson chains representing the bosonic (red) and
    fermionic (blue) bath.
    The dashed boxes indicate the subsystems after each NRG iteration
    starting with the innermost one.
    Adapted from \cite{GlossopIngersent2007}.
  }
\end{figure}

The discretized bath Hamiltonians can be exactly transformed via a Householder
transformation onto two semi-infinite Wilson chains, the fermionic bath
\begin{align}
  \frac{H_{\rm F}}{D} &=  \sum_{\sigma, n=0}^{\infty} \! \!
    \left( \varepsilon_n f_{\sigma,n}^{\dagger} f_{\sigma,n}^{\phantom{\dagger}}
    + \tau_n \left(
      f_{\sigma,n}^{\dagger} f_{\sigma,n+1}^{\phantom{\dagger}} + h.c.
    \right) \right)
\end{align}
and the bosonic bath
\begin{align}
    \frac{H_{\rm B}}{\oc} &= \sum_{m=0}^{\infty}
    \left( \omega_m b_m^{\dagger} b_m^{\phantom{\dagger}} + t_m \left(
      b_m^{\dagger} b_{m+1}^{\phantom{\dagger}} + h.c.
    \right) \right)
    \, .
    \label{eq:13}
\end{align}

By this procedure, the impurity couples only to the first chain link of the
fermionic and the bosonic chain, as it is depicted in Fig.~\ref{fig:wilsonchain}
and $H_0$ is given by
 \begin{align}
  H_0 &= \ed \left( n_{\rm d \uparrow} + n_{\rm d \downarrow} \right)
          + U n_{\rm d \uparrow} n_{\rm d \downarrow}
          + V_0 \sum_{\sigma} \left( f_{0 \sigma}^\dagger d_\sigma
            + d_{\sigma}^\dagger f_{0 \sigma} \right)  \non
      &+ g_0 \left( n_{\rm d} - 1 \right) \left( b_0^{\dagger} + b_0 \right)
\end{align}
with $g_0 = \oc \sqrt{2g/(s+1)}$ \cite{ChengGlossopIngersent2009}.

Since the fermionic bath contains positive and negative energies, whereas the
bosonic bath only contains positive energies, the fermionic Wilson chain
parameters scale as
$\varepsilon_n, \tau_n \propto \Lambda^{-n/2}$,
while the bosonic Wilson chain parameters scale as
$\omega_m, t_m \propto \Lambda^{-m}$.
In order to maintain the correct energy hierarchy in both types of baths,
we need to add a bosonic site only every other NRG iteration as indicated
by the dashed boxes in Fig.~\ref{fig:wilsonchain}.

The Hilbert space of the bosonic operators $b_m$ havs to be restricted within
the numerics to a finite size.
Here, we will use the standard harmonic oscillator eigenbasis,
\ie $b_m^\dagger b_m^{\phantom{\dagger}} \ket{n_m} = n_m \ket{n_m}$,
with $n_m = 0,...,\Nb$ excitations.
The enlarged Hamiltonian is diagonalized by exact diagonalization.
By adding one chain link (or two at each odd iteration) the Hilbert space of
the Hamiltonian grows exponentially with $n$ ($m$).
Therefore, we have to truncate the Hilbert space and keep only the
$\Ns$ eigenstates of the Hamiltonian with the lowest many-body energies.

The renormalization group transformation $R$, $H_{n+1} = R ( H_n )$,
approaches a scale-invariant low-energy fixed point (FP) for large $n$.
As for models with only fermionic baths \cite{Wilson1975}, the FP is
connected to the transformation $R^2$, not $R$ as for bosonic baths,
due to the different ground state spin configurations for chains with
even and odd numbers of sites \cite{GlossopIngersent2007}.

\subsection{The time-dependent numerical renormalization group}
\label{sec:td-nrg}

We use the time-dependent numerical renormalization group (TD-NRG)
\cite{AndersSchiller2005,AndersSchiller2006} to calculate the time evolution
of local impurity quantities after a sudden quench:
$H(t) = \Hi \Theta(-t) + \Hf \Theta(t)$.

Due to the exponential growth of the Hilbert space, high-energy states are
discarded after each iteration.
Since the initial basis set is known, the set of all discarded states forms
a complete basis set and simultaneously serves as an approximate eigenbasis
of the Hamiltonian governing the time evolution of the problem.

Then, the time-dependent expectation value $\expect{O(t)}$ of a
general local operator $\hat{O}$ can be casted into the form
\begin{align}
  \expect{O(t)} = \sum_{n}^{N} \sum_{r,s}^{\rm trun} \;
                    e^{i t (E_{r}^n - E_{s}^n)}
                    O_{r,s}^n \rho^{\rm red}_{s,r}(n)
  \label{eqn:time-evolution-intro}
\end{align}
where $E_{r}^n$ and $E_{s}^n$ are the dimension-full NRG eigenenergies of
the Hamiltonian $\Hf = H(t>0)$ at iteration $n \leq N$ and $O_{r,s}^n$ is
the matrix representation of $\hat{O}$ at that iteration.
$\rho^{\rm red}_{s,r}(n)$ is the reduced density matrix defined as
\begin{align}
  \rho^{\rm red}_{s,r}(n)
    = \sum_{e} \langle s,e;n|\hat{\rho}_{0} |r,e;n \rangle
  \label{eqn:reduced-dm-def}
\end{align}
where $\hat{\rho}_{0}$is the initial density operator of the problem
prior to the quench.
The restricted sum over $r$ and $s$ in Eq.~\eqref{eqn:time-evolution-intro}
requires that at least one of these states is discarded at iteration $n$.

Implementing the TD-NRG requires two NRG runs:
one for the initial Hamiltonian $\Hi = H(t<0)$ to construct the initial
density operator $\hat{\rho}_0$ of the system and one for $\Hf$ to obtain
the approximate eigenbasis governing the time evolution in
Eq.~\eqref{eqn:time-evolution-intro}.
For more details on the TD-NRG see Refs.~\cite{AndersSchiller2005,
AndersSchiller2006}.
Recently, the TD-NRG has been extended to use the full density matrix
including pulsed Hamiltonians \cite{NghiemCosti2014} and periodic switching
\cite{EidelsteinGuettgeSchillerAnders2012}.
Also a hybrid approach \cite{GuettgeAndersSchollwoeckEidelsteinSchiller2013}
with TD-NRG and DMRG extended the scope of application.

In the original implementation of the TD-NRG \cite{AndersSchiller2006}
each phase factor in Eq.~\eqref{eqn:time-evolution-intro}
\begin{align}
   e^{i t (E_{r}^n - E_{s}^n)} \to
  e^{i t (E_{r}^n - E_{s}^n) -\Gamma_n t}
\end{align}
has been supplemented with an energy-resolution dependent damping factor
$\Gamma_n = \alpha \e_n$ proportional to the energy scale
$\e_n = D \Lambda^{-(n-1)/2} ( 1 + 1/\Lambda ) / 2$
at iteration $n$ for all $E_{r}^n - E_{s}^n \not= 0$, and $\alpha = O(1)$.
It mimics an additional decay due to the continuum of modes neglected
by the NRG discretization that is also used for the broadening of the NRG
Lehmann representation of equilibrium spectral functions
\cite{CostiHewsonZlatic1994,PetersPruschkeAnders2006,*WeichselbaumDelft2007,
BullaCostiPruschke2008}.

However, since such an artificial broadening \cite{CostiHewsonZlatic1994}
lacks the detailed information about additional physical decay processes
not included in the NRG Hamiltonian, it could lead to wrongly damp out
oscillatory contributions at long times.
In order to avoid any prejudice, we usually set $\alpha = 0$ and use instead
the  $z$ averaging \cite{YoshidaWithakerOliveira1990,*OliveiraOliveira1994}
to minimize discretization-related oscillations.

\section{Equilibrium properties}
\label{sec:equilibrium-properties}

\subsection{General remarks}

To set the stage for our investigation of the real-time dynamics of the BFAM
in the vicinity to quantum phase transitions, we begin with a summary of the
main equilibrium properties and present the phase diagram of the BFAM in the
parameter space spanned by $U/\Gn$ and $g$.

The equilibrium properties of the BFAM are well understood by
several publications \cite{BullaPruschkeHewson1997,GonzalezIngersent1998,
BullaGlossopLoganPruschke2000,IngersentSi2002,VojtaFritz2004,
FritzVojta2004,FritzVojta2013,GlossopKhoshkhouIngersent2008,
ChengGlossopIngersent2009,PixleyKirchnerIngersentSi2013}.
Due to the richness of the phase diagram of the  BFAM we give an overview
of the different NRG fixed points (FPs) and explain by which quantities
we classify the phases.

Within this paper, we focus on a particle-hole (ph) symmetric final
Hamiltonian $\Hf$, \ie $\Uf = - 2 \edf$.
We will investigate two classes of quenches:
(i) $\Hi$ also obeys ph symmetry and (ii) $\Hi$ describes an
initially ph symmetry broken phase.

In the first case, the level occupancy is $\ndt = 1$ for all times,
and there are two pairs of locally degenerated states:
the empty- and the double-occupied state $\ket{0},\ket{2}$ and,
in absence of an external magnetic field the two spin states
$\ket{\sigma}$ which contain exactly one electron.

Depending on the combination matrix $(r,s)$ three types of quantum
critical regimes have been identified in the model, see Fig.~13 in
Ref.~\cite{PixleyKirchnerIngersentSi2013}.
For small $r$, the bosonic bath properties govern the critical exponents
of the thermodynamic properties defining a B-type regime with a boundary
$0.5 \leq s = 1 - 2 r \leq 1$.
For larger $r$ and $s$ the critical exponents are given by the fermionic
bath properties, hence defining the F-type regime with a non-linear boundary.
Between the two regimes there is a mixed (M) type regime where
the exponents do not decompose in fermionic and bosonic parts.

Firstly, we show below, that the real-time dynamics in the F-type regime
can essentially be reproduced by a pseudo-gap SIAM with a
renormalized $\Uren$ while the non-equilibrium dynamics in the
B-type regime differs from those of an effective SIAM and are affected
by an additional damping provided by the relevant bosonic modes present in the
fixed point spectrum.

Secondly, starting from an initially ph symmetry broken state,
we track the real-time dynamics of the level occupancy $\ndt$,
but restrict ourselves to $r=0$.
A thorough discussion of the corresponding equilibrium properties of the
asymmetric model can be found in Refs.~\cite{ChengGlossopIngersent2009,
PixleyKirchnerIngersentSi2013} and are only relevant for the initial
preparation of the system and entering $\rho_0$.

For the calculations of the BFAM we use $U/\Gn=1$, $\Gn/D=0.01$ and
vary over the bosonic coupling $g$.
For the comparison with the SIAM, we also use $\Gn/D=0.01$ but vary
over $U/\Gn$ with the same fermionic bath exponent $r$.
All our (TD-)NRG calculations are done with $\Lambda = 6$, and $N = 40$
NRG iterations.
For the BFAM we use $\Ns = 1000$, $\Nb = 10$, while for the SIAM we keep
instead $\Ns = 2000$ states at each iteration.
We average over $\Nz = 8$ different bath realizations.
For the equilibrium results we forgo sometimes this averaging
(no $z$ averaging) to minimize the numerical effort.
Therefore, the numerical values of the critical couplings depend on
whether or not the $z$ averaging has been done.

\subsection{Renormalization of the Coulomb repulsion}
\label{sec:eq-renormalized-coulomb-repulsion}

In order to understand qualitatively the major features of the phase diagram
and construct an effective SIAM, we briefly review the boson-mediated
attractive electron-electron interaction.
To keep it simple, we consider a decoupled fermionic bath by setting $\Gn=0$.
Then, the remaining Hamiltonian reads
\begin{align}
  \label{eq:H-coulomb-repulsion}
  H^\prime
    &= \epsilon_{\rm d} \left( n_{\rm d \uparrow} + n_{\rm d \downarrow} \right)
       + U n_{\rm d \uparrow} n_{\rm d \downarrow}
       + \sum_{\vec{q}} \omega_{\vec{q}} b_{\vec{q}}^\dagger
           b_{\vec{q}}^{\phantom{\dagger}} \non
    &\quad + \left( \nd - 1 \right) \sum_{\vec{q}} g_{\vec{q}}
       \left( b_{\vec{q}}^\dagger + b_{\vec{q}}^{\phantom{\dagger}} \right)
  \quad .
\end{align}

Since the number of fermions is conserved, $H^\prime$ can be diagonalized
in each particle number subspace separately.
For $\ket{\sigma}$ the fermion-boson interaction vanishes
and the bath decouples.
The two other states, $\ket{0}$ and $\ket{2}$, require displaced harmonic
oscillator operators $\tilde b_{\vec{q}} = b_{\vec{q}} + \theta_{\vec{q}}$
and the effective bath Hamiltonian is given by \cite{ChengGlossopIngersent2009}
\begin{align}
  \braket{0}{H}{0} &= \sum_{\vec{q}} \omega_{\vec{q}}
      \tilde b_{\vec{q}}^\dagger \tilde b_{\vec{q}}^{\phantom{\dagger}}
    - \sum_{\vec{q}} \frac{g_{\vec{q}}^2}{\omega_{\vec{q}}} \qquad \text{and} \\
  \braket{\uparrow\downarrow}{H}{\uparrow\downarrow} &= 2 \ed + U
    + \sum_{\vec{q}} \omega_{\vec{q}} \tilde b_{\vec{q}}^\dagger
        \tilde b_{\vec{q}}^{\phantom{\dagger}}
    - \sum_{\vec{q}} \frac{g_{\vec{q}}^2}{\omega_{\vec{q}}}
    \non
    &= 2 \ed + \Uren
    + \sum_{\vec{q}} \omega_{\vec{q}} \tilde b_{\vec{q}}^\dagger
        \tilde b_{\vec{q}}^{\phantom{\dagger}}
\end{align}
with the displacements $\theta_{\vec{q}} = \pm g_{\vec{q}} / \omega_{\vec{q}}$ 
and the renormalized Coulomb repulsion
\begin{align}
  \label{eq:uren}
  \Uren = U - \sum_{\vec{q}} \frac{g_{\vec{q}}^2}{\omega_{\vec{q}}}
        = U - g \frac{2 \oc}{s}
\end{align}
using the bosonic coupling function defined in Eqs.~\eqref{eq:J-w-def} and
\eqref{eq:J-w-pl}.

This effective Coulomb repulsion changes sign and becomes attractive
for $g > U s / (2 \oc)$.
The sign change causes a level crossing between the two degenerated
magnetic states $\ket{\sigma}$ and the degenerated charge states
$\ket{0},\ket{2}$ for ph-symmetric parameters.
At $\gp = U s/(2 \oc)$ the renormalized Coulomb repulsion vanishes:
$\Uren(\gp) = 0$.

\subsection{Definition of the effective spin and charge moments}
\label{sec:eq-def-delta-X}

The effective local moment \cite{Wilson75}
\begin{align}
  \mueff(T)
    = \Delta \left( \expect{S_{\rm z}^2 (T)} - \expect{S_{\rm z}(T)}^2 \right)
  \label{eq:mueff}
\end{align}
can be used to identify the  quantum critical point (QCP).
$\Delta(X)$ measures the macroscopic observable $X$ in presence of the
impurity and the Wilson chain and subtracts the effect of the quantity
$X$ of the pure Wilson chain without the impurity at temperature $T$,
and $S_{\rm z}$ denotes the z component of the total spin of the system.

In the continuum $\expect{X},\expect{X^2}$ are extensive and diverge
with the system size.
By the definition of $\Delta(X)$ \cite{Wilson1975,GonzalezIngersent1998}
the impurity contribution of the quantity $X$ is extracted, which is of
the order $O(1)$.
Note, however, that $\mueff(T)$ does not measure the local impurity spin
observable but rather the difference in the total system properties with
and without the impurity.
Hence, the effective impurity spin moment $\mueff(T)$ is in general related
to a degree of freedom (DOF) comprising a linear combination of local and
conduction electron spin observables.

Analog to $\mueff(T)$, we can define an effective charge moment
\begin{align}
  \Qeff(T)
    = \frac{1}{4} \Delta \left( \expect{Q^2 (T)} - \expect{Q(T)}^2 \right)
  \label{eq:chic}
\end{align}
whereby $Q$ is the total charge of the system with respect to half-filling.

\subsection{Overview of the phases}
\label{sec:eq-overview-phases}

\begin{figure}[b]
  \includegraphics[width=\linewidth]{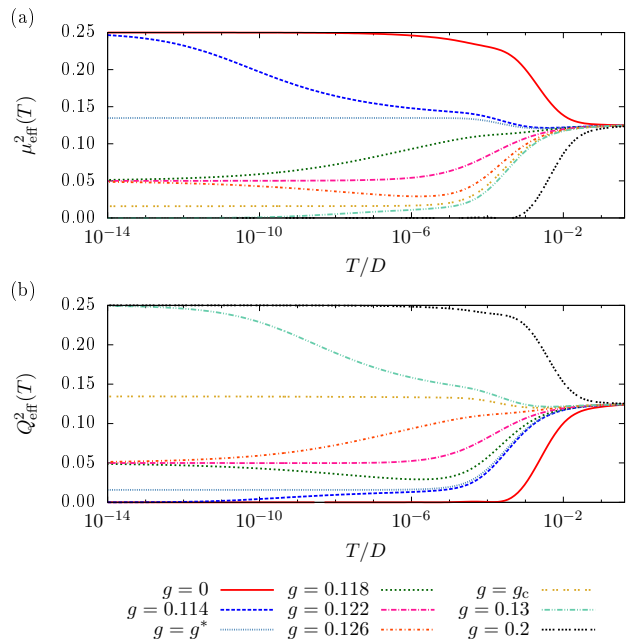}
  \caption{\label{fig:equilibrium}%
    (Color online)
    Effective moments (a) $\mueff(T)$ and (b) $\Qeff(T)$ for bath exponents
    $(r,s)=(0.4,0.8)$ without $z$ averaging.
    The BFAM parameters are $\Gn/D = 0.01$ with the ratio $U/\Gn = 1$.
    Different FPs are reached for $T \to 0$.
    The unstable critical FPs have the couplings $\gs=0.11496(5)$ and
    $\gc=0.12847(7)$.
  }
\end{figure}

Ignoring the details of the F-, B- and M-type regimes, three stable FPs
are found for the ph-symmetric BFAM for $r > 0$, $1/2 < s < 1$ and a
finite $U > \Uc$.

Starting with $g=0$ in the local moment (LM) FP, an effective spin $S = 1/2$
DOF decouples from the problem as it is indicated by
$\lim_{T \to 0} \mueff(T) = 1/4$ and $\lim_{T \to 0} \Qeff(T) = 0$.

Increasing $g$ reduces $\Uren$ such that the effective Kondo interaction
exceeds the critical value:
In the Kondo-screened symmetric strong coupling (SSC) FP the impurity
spin is partially screened with $\lim_{T \to 0} \mueff(T) = r/8$ and
simultaneously $\lim_{T \to 0} \Qeff(T) = r/8$.
This universality of the effective moments indicates that the conventional
spin Kondo effect and the charge Kondo effect are adiabatically connected
and can only the separated by their crossover behavior:
There exists only one universal FP in the ph-symmetric regime.
In the spin Kondo regime, $\mueff(T) $ approaches the value $r/8$ from above
and $\Qeff(T)$ from below when lowering the temperature, as seen in
Fig.~\ref{fig:equilibrium}.
In the charge Kondo regime the role of $\Qeff(T)$ and $\mueff(T)$ is
interchanged.
The transition from the LM to the SSC FP occurs at $g = \gs$ and is governed
by an unstable FP.

Once $g$ exceeds $\gc > g^*$ with $\Uren < 0$, we reach the localized (L) FP
characterized by a free charge moment, \ie $\lim_{T \to 0} \Qeff(T) = 1/4$
and $\lim_{T \to 0} \mueff(T) = 0$.
Since a free charge moment decouples from the continuum and the spin momentum
vanishes, this FP mirrors the LM FP where $\Uren > 0$.

The two QCPs at $g = \gc$ and $g = \gs$, governing the transition between the
LM FP and the SSC FP ($\gs$) and the transition between the SSC FP and the L FP
($\gc$), are characterized by two unstable FPs with
$\Qeff(T \to 0, \gc)= \mueff(T \to 0, \gs)$ and
$\Qeff(T \to 0, \gs)= \mueff(T \to 0, \gc)$.

This series of FPs can be  reached by a single parameter scan as long as
$U > \Uc$ for $g=0$.
$\Uc$ separates the LM FP and an SSC FP in the pg SIAM in the absence of
a bosonic bath \cite{BullaPruschkeHewson1997,GonzalezIngersent1998}.
For $0 < U < \Uc$, the LM FP is absent and only the SSC FP remains for $g=0$.

\begin{figure}[tb]
  \includegraphics[width=0.9\linewidth]{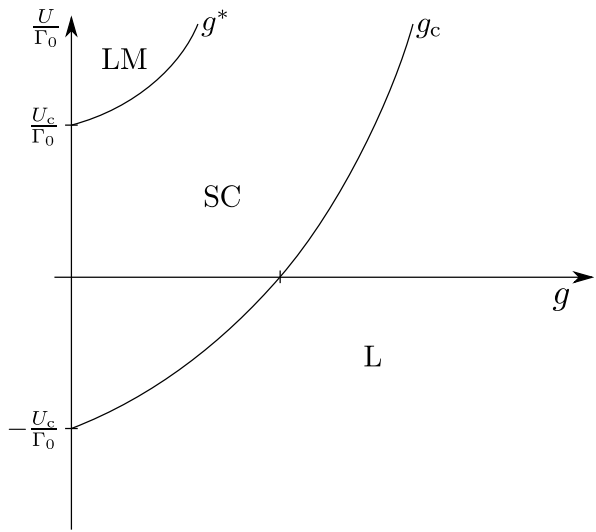}
  \caption{\label{fig:phasediagram}
    Sketch of the phase diagram of the BFAM in dependence of the
    ratio $U/\Gn$ and the bosonic coupling $g$.
    The bosonic bath exponent is sub-ohmic $1/2 < s < 1$ and the fermionic
    exponent is $0 < r < 1/2$.
    The critical $\Uc/\Gn$ is already known for the SIAM.
    The phase boundaries $\gs$ and $\gc$ separate the local moment (LM), the
    symmetric strong coupling (SSC) and the localized (L) phase, respectively.
  }
\end{figure}

We illustrate the approach of the different FPs for a F-type exponent
combination $(r,s)=(0.4,0.8)$ and $U/\Gn=1$, $\Gn/D=0.01$ in
Fig.~\ref{fig:equilibrium}.
$\mueff(T)$  is depicted in Fig.~\ref{fig:equilibrium}(a) and $\Qeff(T)$
is shown in Fig.~\ref{fig:equilibrium}(b) for the same set of bosonic
couplings $g$.
All 5 FP values (3 stable, 2 unstable) emerge in the zero-temperature limit
by varying over the bosonic coupling $g$.

We complete our overview with the generic phase diagram for $0 < r < 1/2$ and
$1/2 < s < 1$ shown in Fig.~\ref{fig:phasediagram} where we have extended the
previous investigations \cite{ChengGlossopIngersent2009,
PixleyKirchnerIngersentSi2013} to negative $U$.
The phase boundaries $(U,\gs)$ and $(U,\gc)$ start at the well-known phase
boundary points of the SIAM \cite{BullaPruschkeHewson1997,FritzVojta2013}:
$\Uc/\Gn$ separates the LM and the SSC phases and $-\Uc/\Gn$ separates
the SSC and L phases.
For infinitely large $U/\Gn$ both phase boundaries $\gs$ and $\gc$ merge.

It contains an important message:
The three stable FPs can either be reached for fixed $U > \Uc$ by tuning $g$
or for a fixed $g$ by tuning $U$.
As discussed in Sec.~\ref{sec:eq-renormalized-coulomb-repulsion} above,
the leading order effect of the bosonic bath coupling is the reduction of
$U$ to a renormalized value which is given by Eq.~\eqref{eq:uren} in the
absence of the fermionic bath.

\subsection{Equilibrium double occupancy}
\label{sec:eq-double-occupancy}

\begin{figure}[b]
  \includegraphics[width=\linewidth]{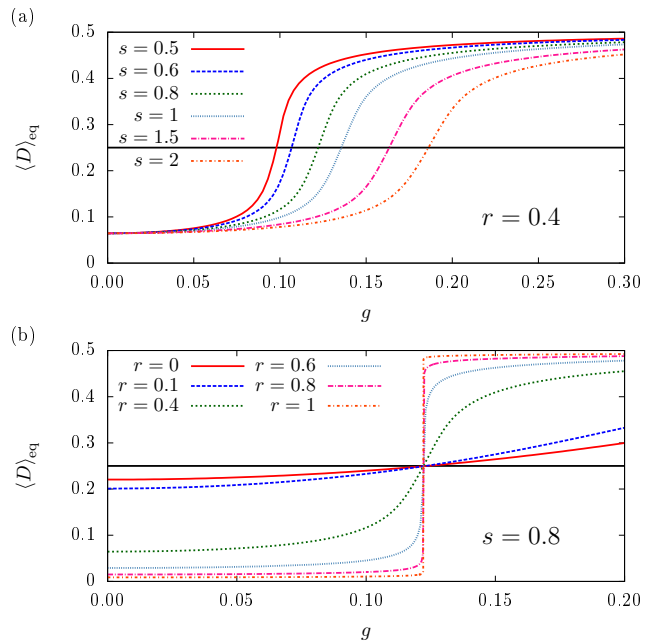}
  \caption{\label{fig:Deq-bfam-r-s-var}%
    (Color online)
    Equilibrium double occupancy $\Deq$ of the BFAM versus the bosonic
    coupling $g$ for the fixed ratio $U/\Gn = 1$, $\Gn/D = 0.01$ and without
    $z$ averaging.
    All data is obtained in the limit $T \to 0$.
    Varying (a) over $s$ for fixed $r=0.4$ and (b) over $r$ for fixed $s=0.8$.
  }
\end{figure}

Now we turn to the equilibrium double occupancy $\Deq$ for different bath
exponent combinations $(r,s)$.
While $\mueff(T \to 0)$ and $\Qeff(T \to 0)$ jump at and across the QCP,
$\Deq$ varies continuously \cite{GonzalezIngersent1998,KleineMusshoffAnders2014}
as function of $g$ as demonstrated in Fig.~\ref{fig:Deq-bfam-r-s-var} for
$U/\Gn=1$ and $\Gn/D=0.01$.

Although the estimate in Eq.~\eqref{eq:uren} for $\Uren$ does not hold at
finite $\Gn$, we can generalize the definition of $\gp$:
\begin{align}
  \Deq \! (\gp) = 1/4
  \quad .
\end{align}
At this value the occupation probability of all four local states are equal
corresponding to $U = 0$ for $g = 0$ and the value is indicated by the solid
horizontal line in Fig.~\ref{fig:Deq-bfam-r-s-var}.
As discussed above, $\gp$ can be viewed as the crossover parameter from the
spin Kondo to the charge Kondo regime in the interval $\gs < \gp < \gc$.
Since $\Uren$ is essentially governed by the bosonic bath, all curves in
Fig.~\ref{fig:Deq-bfam-r-s-var}(b) cross each other in one point,
$\Deq = 1/4$, revealing that $\Uren(g,s)=0$ is independent of $r$.

\begin{figure}[t]
  \includegraphics[width=\linewidth]{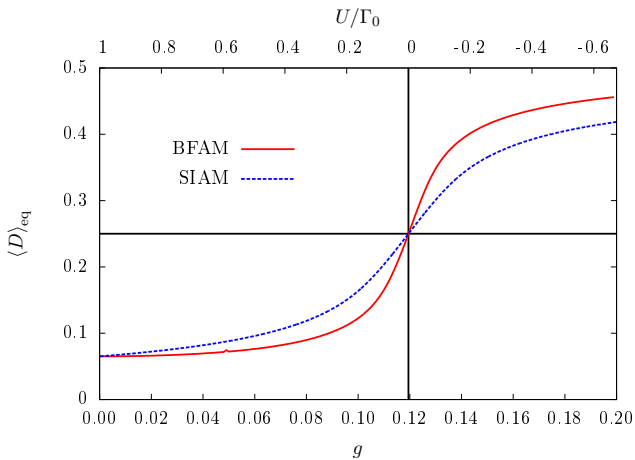}
  \caption{\label{fig:Deq-bfam-siam-I}%
    (Color online)
    Equilibrium double occupancy $\Deq$ of the BFAM versus
    the bosonic coupling $g$ for the fixed ratio $U/\Gn = 1$, $\Gn/D=0.01$,
    $(r,s)=(0.4,0.8)$ and of the SIAM versus the Coulomb interaction $U/\Gn$
    for fixed $\Gn/D=0.01$ and $r=0.4$.
    All data is obtained in the limit $T \to 0$ and averaged over $\Nz=8$
    realizations.
  }
\end{figure}

Since the three stable phases of the BFAM can be reached for $g=0$ as well
by tuning $U$, the question arises what is the influence of the additional
bosonic bath beyond a renormalization $U \to \Uren(g)$ on the real-time
dynamics compared to the SIAM.
$\Deq$ monotonically increases with increasing $g$ in the BFAM and decreases
with increasing $U$ in the SIAM, as seen in Fig.~\ref{fig:Deq-bfam-siam-I}.
By the requirement
\begin{align}
  \label{eq:requirement}
  \Deq^{\rm BFAM}(g) = \Deq^{\rm SIAM}(U/\Gn)
\end{align}
we define an effective SIAM that leads to the same local fermionic
equilibrium expectation values for the same $r$.

\section{Real-time dynamics for the symmetric model}
\label{sec:td-nrg-results}

\subsection{F-type bath exponents}
\label{sec:td-nrg-f-type}

\begin{table}[t]
  \caption{\label{tab:g-U}%
    Correspondence between the bosonic coupling $g$ of the BFAM with fixed
    ratio $U/\Gn = 1$, $\Gn/D=0.01$, $(r,s)=(0.4,0.8)$ and the Coulomb
    interaction $U/\Gn$ of the SIAM defined via Eq.~\eqref{eq:requirement}.
    All data is obtained in the limit $T \to 0$.
  }
  \begin{ruledtabular}
    \begin{tabular}{lccr}
       Phase  &      $\Deq$    &  $g$           &  $U/\Gn$     \\ \colrule
      \bf LM  &  \bf 0.065119  &  \bf 0         &  \bf 1       \\
              &      0.069507  &  0.04          &  0.8917      \\
              &      0.076314  &  0.06          &  0.7557      \\
              &      0.089795  &  0.08          &  \bf 0.5628  \\
              &      0.122573  &  0.1           &  0.3098      \\ \colrule
              &      0.195940  &  0.114         &  0.09381     \\
              &      0.214397  &  0.116         &  0.06031     \\
              &      0.234426  &  0.118         &  0.02601     \\
     \bf SSC  &  \bf 0.250000  &  \bf 0.119488  &  \bf 0       \\
              &      0.276720  &  0.122         &  -0.0449     \\
              &      0.297465  &  0.124         &  -0.0816     \\
              &      0.316741  &  0.126         &  -0.1188     \\ \colrule
              &      0.392307  &  0.14          &  -0.3999     \\
              &      0.428959  &  0.16          &  -0.8580     \\
              &      0.446127  &  0.18          &  -1.3860     \\
       \bf L  &  \bf 0.456499  &  \bf 0.2       &  \bf -1.9872
    \end{tabular}
  \end{ruledtabular}
\end{table}

As an example for the generic real-time dynamics in the BFAM we present
quenches for fixed $U/\Gn=1$, $\Gn/D=0.01$, $(r,s)=(0.4,0.8)$ by a sudden
change of the coupling constant $g$ from its initial value $\gi$ to the
final value $\gf =: g$.

We have picked three representative initial values:
(i) $\gi = 0$ starting from the LM FP,
(ii) $\gs < \gi = 0.119488 < \gc$ corresponding to $\Deq\!(\gi) = 1/4$
($\Uren= 0)$ starting from the SSC FP, and
(iii) $\gc < \gi = 0.2$ deep in the L regime.
For $\gf=:g$ we select a series of values within each phase.

We will show that for the F-type bath exponents the real-time dynamics can
be fully understood by mapping the BFAM onto an effective SIAM defined
by Eq.~\eqref{eq:requirement}.
The corresponding values of $U/\Gn$ for the given $g$ can be found in
Tab.~\ref{tab:g-U}.
Hence, the quenches in $g$ within the BFAM are mapped onto the corresponding
$U$ quenches in the SIAM.

\subsubsection{Quenches within a phase}
\label{sec:quenches-within}

\begin{figure}[tb]
  \includegraphics[width=\linewidth]{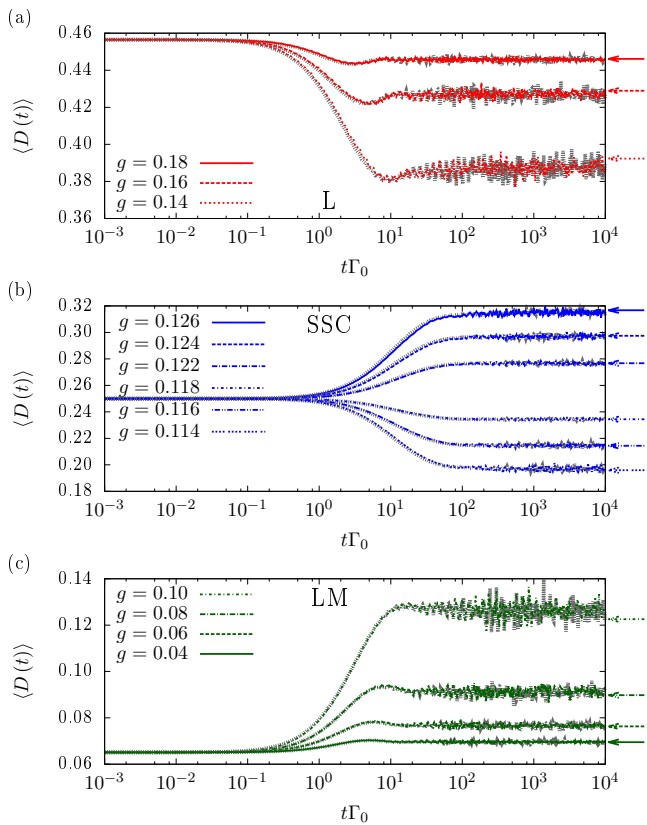}
  \caption{\label{fig:Dt-F-within}%
    (Color online)
    Real-time dynamics of the double occupancy $\Dt$ (colored curves) in
    the BFAM for bath exponents $(r,s)=(0.4,0.8)$ with $U/\Gn=1$ and
    $\Gn/D=0.01$.
    Gray curves in the background are the effective dynamics of the
    double occupancy in the SIAM.
    We perform quenches within the (a) L phase, (b) SSC phase and (c) LM phase.
    Details on the chosen couplings $\gi$ and $g$ are given in the beginning
    of Sec.~\ref{sec:td-nrg-f-type} as well as in Tab.~\ref{tab:g-U}.
  }
\end{figure}

In Fig.~\ref{fig:Dt-F-within} we present the quenches within each of the
three stable phases.
We supplement the dynamics of the BFAM with those of the effective SIAM
plotted in gray (color online) in the background.
For all those quenches we find equilibration and the dynamics of the BFAM
agrees with the effective SIAM within the numerical accuracy of the TD-NRG.

The quenches within the L phase are shown in Fig.~\ref{fig:Dt-F-within}(a)
where we start with a large negative $\Uren$ ($\gi=0.2$) and $\expect{D(t=0)}$
close to $1/2$.
A sudden reduction of $g$ results in a reduction of the double occupancy.
The double occupancy $\Dt$ smoothly declines on a timescale roughly given
by $1/\Gn$ with some small $g$-dependent corrections.

Fig.~\ref{fig:Dt-F-within}(b) depicts the data for quenches starting from
the degeneracy point of all local states ($\Uren=0$) using $\gi = 0.119488$.
For $g > \gi$, $\Dt$ increases, while for $g < \gi$, $\Dt$ decreases with time.
The corresponding thermal expectation values, indicated as arrows in the figure,
demonstrate thermalization in the long-time limit.

Starting from a decoupled bosonic bath ($\gi=0$), quenches within the LM phase
are shown in Fig.~\ref{fig:Dt-F-within}(c).
Similar to Fig.~\ref{fig:Dt-F-within}(a) the equilibrated value of $\expect{D}$
deviates from the thermodynamic equilibrium:
the larger $|\gf - \gi|$ the larger the deviation.
This expected behavior is related to the decoupling of a charge DOF from
the bath continuum in the L FP.
Note, however, that the deviations are very small.

In addition, we have investigated the same type of quenches using an effective
SIAM, by setting $\gi = \gf = 0$ and switching between the corresponding $U$
values that lead to identical $\Deq$.
The real-time dynamics of $\Dt$ has been added as gray curves to the graphs of
Fig.~\ref{fig:Dt-F-within}.
For all investigated cases, the effective SIAM results agree perfectly
with those of the BFAM demonstrating that the effect of the bosonic bath for
F-type bath exponents is fully determined by $\Uren$ and the fermionic bath
exponent $r$.

\begin{figure}[tb]
  \includegraphics[width=\linewidth]{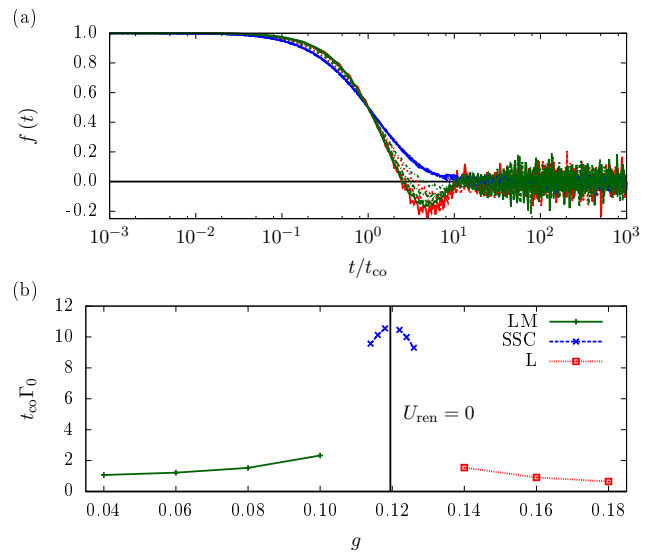}
  \caption{\label{fig:Dt-F-within-rescaled}%
    (Color online)
    (a) Scaled dynamics of all BFAM quenches in Fig.~\ref{fig:Dt-F-within}
    according to Eq.~\eqref{eq:ft} versus the crossover time scale $\tco$.
    (b) Crossover time scale $\tco$ versus $g$.
  }
\end{figure}

In order to extract the relevant crossover time scale $\tco$, we introduce
the scaling function \cite{AndersSchiller2006,KleineMusshoffAnders2014}
\begin{align}
  f(t) = \frac{\Dt - \expect{D(\infty)}}{\expect{D(0)} - \expect{D(\infty)}}
  \label{eq:ft}
\end{align}
of the time-dependent double occupancy $\Dt$.
The function maps the dynamics onto $f(0) = 1$ and approaches $f(\infty) = 0$
at infinitely long times independent of the parameters.
We define the crossover time scale $\tco$ by demanding $f(\tco) = 1/2$ and
plot $f(t)$ versus $t/\tco$ in Fig.~\ref{fig:Dt-F-within-rescaled}(a) using
the data of Fig.~\ref{fig:Dt-F-within}.
Universality of the quenches in the SSC phase (blue curves) is clearly visible.
For the L phase and the LM phase comparable shapes of $f(t/\tco)$ are found,
which slightly differ from the SSC case and are non-universal for $t > \tco$.

The extracted $\tco$ versus $g$ is shown in
Fig.~\ref{fig:Dt-F-within-rescaled}(b) for the three different phases.
The vertical line marks the coupling $\Deq(\gp=0.119488) = 1/4$
corresponding to $\Uren = 0$. 
The crossover time scale shows an ascending slope for $\Uren \to 0^+$ and
a descending slope for an increasing attractive $\Uren \to -\infty$
independent of the quench type.
For the quenches using small values of $U$ as reported in
Ref.~\cite{KleineMusshoffAnders2014}, the variation of $\tco$ is of order of
3\% which we had attributed to numerical errors at that time.
Therefore, we have claimed in Ref.~\cite{KleineMusshoffAnders2014} that
$\tco$ remains independent of $U$.
Our present study \footnote{%
  We like to note that the absolute value of $\tco$ in our present study
  is roughly a factor of two greater compared to the inset of Fig.~4(b)
  in Ref.~\cite{KleineMusshoffAnders2014}.
  This originates from the different value of $\Lambda$ used in the present
  paper since we did not include the discretization correction factor
  $A(\Lambda)$ in our calculation \cite{KrishnaWilkinsWilson1980,
  KrishnaWilkinsWilson1980b}.
}, however, suggests a dependence $\tco(\Uren)$ as depicted in
Fig.~\ref{fig:Dt-F-within-rescaled}(b).

\subsubsection{Quenches from the SSC FP}
\label{sec:quenches-from-sc}

\begin{figure}[t]
  \includegraphics[width=\linewidth]{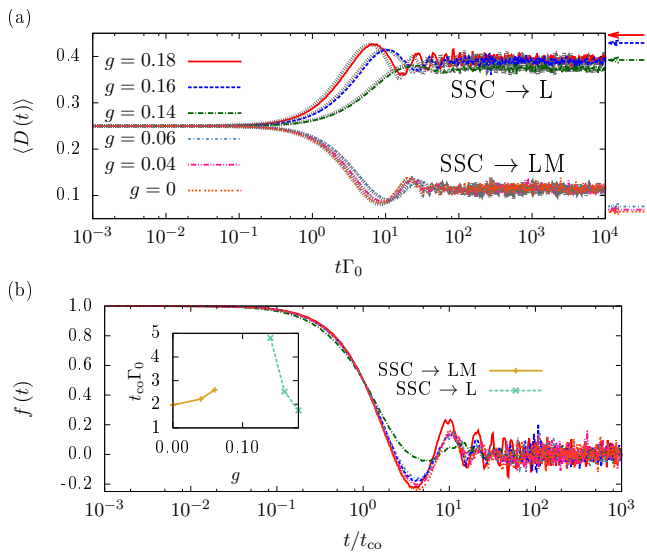}
  \caption{\label{fig:Dt-F-fromSC}%
    (Color online)
    Real-time dynamics of the double occupancy $\Dt$ (colored curves) in
    the BFAM with the same parameters as in Fig.~\ref{fig:Dt-F-within}.
    Gray curves in the background show the corresponding double occupancy
    in the SIAM.
    (a) Quenches with the initial condition $\gi=0.119488$ (SSC FP) into the
    L and LM phase.
    (b) Scaled data of BFAM quenches of (a) via Eq.~\eqref{eq:ft} versus
    the time scale $\tco$, which is depicted in the inset.
  }
\end{figure}

In this section, we focus on quenches from the SSC FP with
$\expect{D(\gi=0.119488)} = 1/4$ into the L and LM phase.
We have found damped oscillations for both cases which equilibrate to a
steady-state values significantly deviating from the thermodynamic
equilibrium as shown in Fig.~\ref{fig:Dt-F-fromSC}(a).
The arrows at the right side of Fig.~\ref{fig:Dt-F-fromSC}(a) illustrate the
equilibrium double occupancy calculated by an independent NRG calculation for
the final conditions.
The oscillation frequency depends on the renormalized Coulomb repulsion
$|\Uren|$ whose values can be estimated from the effective SIAM as listed
in Tab.~\ref{tab:g-U}.

The stronger $g$ deviates from the critical coupling $\gs$ and $\gc$
respectively, the more the effective local charge or spin moments are
localized and contribute to $\Dt$:
the damping is reduced leading to more pronounced oscillations.

We have extracted the crossover time scale $\tco$ by plotting $f(t)$ using
the data of Fig.~\ref{fig:Dt-F-fromSC}(a) in Fig.~\ref{fig:Dt-F-fromSC}(b)
as function of $t/\tco$.
Now, the locations of the oscillatory maxima and minima agree very well.
The inset of Fig.~\ref{fig:Dt-F-fromSC}(b) shows the dependency of $\tco$
on $g$ translating to $\tco \approx 2/|\Uren|$.

\subsubsection{Quenches into the SSC phase}
\label{sec:quenches-into-sc}

\begin{figure}[t]
  \includegraphics[width=\linewidth]{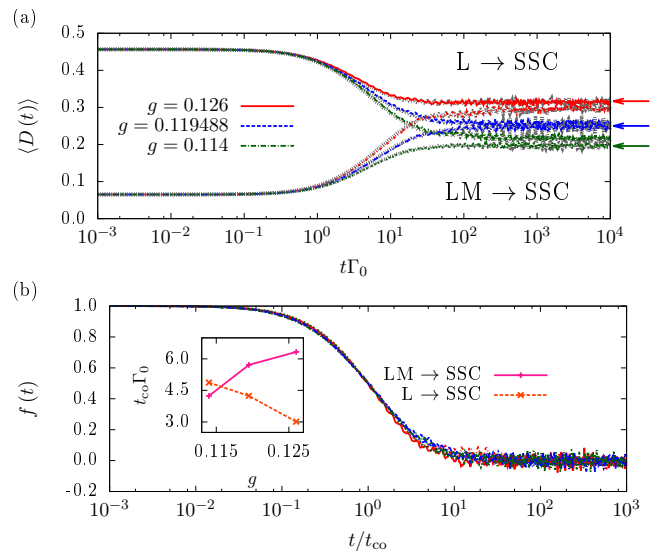}
  \caption{\label{fig:Dt-F-toSC}%
    (Color online)
    Real-time dynamics of the double occupancy $\Dt$ (colored curves) in
    the BFAM with the same parameters as in Fig.~\ref{fig:Dt-F-within}.
    Gray curves in the background show the corresponding double occupancy
    in the SIAM.
    Quenches from the L and LM starting points into the SSC phase.
    (a) $\Dt$ versus $t\Gn$.
    (b) Data of BFAM quenches of (a) scaled via Eq.~\eqref{eq:ft} versus $\tco$.
    Inset in (b) shows dependency of $\tco$ on $g$.
  }
\end{figure}

Now we reverse the quench direction compared to the previous section:
We start from the LM or L FP and quench into the SSC phase.
For both quench types the double occupancy $\Dt$ equilibrates
onto a steady-state value via a smooth crossover as demonstrated
in Fig.~\ref{fig:Dt-F-toSC}.

However, the steady-state values for the same $\Hf$ are slightly different:
The value for a quench starting from the L FP always exceeds the value for
the quench starting from the LM FP.
Nevertheless, the deviation from the equilibrium NRG value remains very small
as indicated by arrows on the right side of Fig.~\ref{fig:Dt-F-toSC}(a) and
is related the TD-NRG discretization errors.
Therefore, we conclude that the quenches into the SSC phase thermalize in
the continuum limit.

Fig.~\ref{fig:Dt-F-toSC}(b) reveals perfect universality of the function
$f(t/\tco)$.
This illustrates that the quenches from the L or LM FP are mirror images
of each other.

\subsubsection{Quenches between the LM and L phase}
\label{sec:quenches-between-lml}

\begin{figure}[t]
  \includegraphics[width=\linewidth]{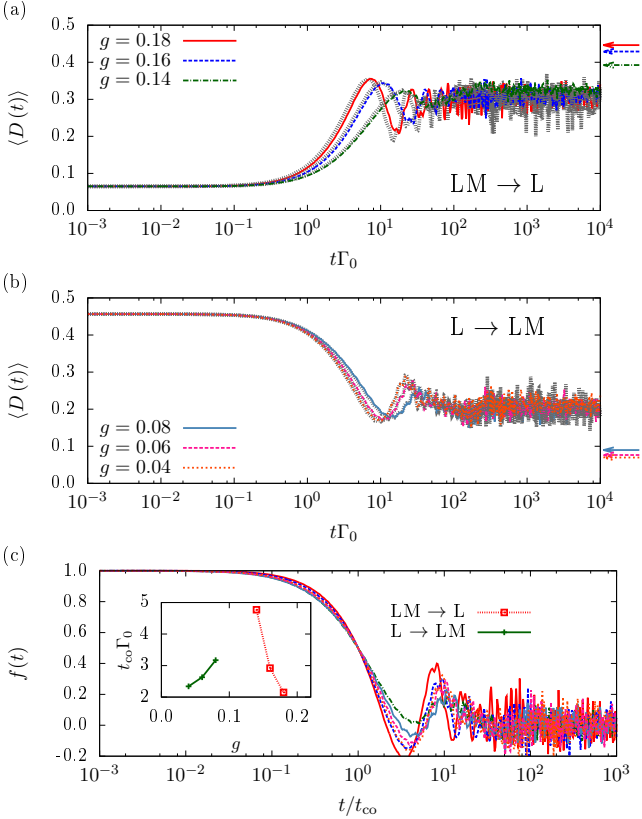}
  \caption{\label{fig:Dt-F-LMvsL}%
    (Color online)
    Real-time dynamics of the double occupancy $\Dt$ (colored curves) in
    the BFAM with the same parameters as in Fig.~\ref{fig:Dt-F-within}.
    Gray curves in the background show the corresponding double occupancy
    in the SIAM.
    (a) Quenches from the LM starting point into the L phase and
    vice versa in (b).
    (c) Scaled dynamics of all BFAM quenches via Eq.~\eqref{eq:ft} versus
    the time scale $\tco$.
    Inset in (c) shows dependency of $\tco$ on $g$.
  }
\end{figure}

Starting from the LM phase and quenching into the L phase or vice versa
are the two extremest types of quenches in the model.
The results are depicted in Fig.~\ref{fig:Dt-F-LMvsL}(a) and
Fig.~\ref{fig:Dt-F-LMvsL}(b) respectively.
A steady-state value is approached which differs significantly from the
equilibrium value and even stronger compared to quenching over one QCP.
$\tco$ is shown in the inset of Fig.~\ref{fig:Dt-F-LMvsL}(c) and is very
similar in absolute value and functional dependency on $g$ as in
Fig.~\ref{fig:Dt-F-fromSC}.

\subsubsection{Short summary for the dynamics for F-type bath exponents}
\label{sec:summary-F}

Our detailed analysis of the real-time dynamics with F-type bath exponents
has demonstrated equilibration towards steady-state values which differ from
the thermal equilibrium if the system is quenched into the L or the LM phase.
We have also investigated the non-equilibrium dynamics using an effective SIAM
defined by $\Uren$ and could show that its dynamics reproduce exactly
the one of the more complicated BFAM as exemplified in
Fig.~\ref{fig:Dt-F-within}.

\subsection{B-type bath exponents}
\label{sec:quenches-bosonic-type}

\begin{table}[t]
  \caption{\label{tab:g-U-B}%
    Correspondence between the bosonic coupling $g$ of the B-type BFAM and
    the Coulomb repulsion $U/\Gn$ of the SIAM with respect to
    Eq.~\eqref{eq:requirement} and model parameters as described in the
    beginning of Sec.~\ref{sec:quenches-bosonic-type}.
  }
  \begin{ruledtabular}
    \begin{tabular}{lccr}
        Phase  &      $\Deq$      &      $g$    &     $U/\Gn$    \\ \colrule
        \bf L  &  \bf 0.48729475  &  \bf 0.50   &  \bf -30.5572  \\
   (L in SIAM) &      0.48601585  &      0.48   &      -27.7935  \\
               &      0.48455407  &      0.46   &      -25.1727  \\
               &      0.48287547  &      0.44   &      -22.6981  \\
               &      0.48093136  &      0.42   &      -20.3643  \\ \colrule
        \bf L  &  \bf 0.434957    &  \bf 0.25   &  \bf  -5.98    \\
 (SSC in SIAM) &      0.427673    &      0.24   &       -5.43    \\
               &      0.418887    &      0.23   &       -4.84    \\
               &      0.407986    &      0.22   &       -4.26    \\ \colrule
               &      0.276267    &      0.13   &       -0.533   \\
               &      0.264578    &      0.12   &       -0.295   \\
               &      0.253931    &      0.11   &       -0.079   \\
      \bf SSC  &  \bf 0.25        &  \bf 0.106  &  \bf   0       \\
 (SSC in SIAM) &      0.244341    &      0.10   &        0.115   \\
               &      0.235799    &      0.09   &        0.286   \\
               &      0.228286    &      0.08   &        0.439
    \end{tabular}
  \end{ruledtabular}
\end{table}

For bath exponents of the B-type $(r,s)=(0.1,0.6)$ the BFAM exhibits the
same three phases but the critical exponents of the equilibrium thermodynamics
are determined by the properties of the bosonic bath
\cite{PixleyKirchnerIngersentSi2013}.
The question arises whether the non-equilibrium dynamics is also governed by
the low-energy modes of bosonic bath or whether only an $\Uren$ is generated
and the dynamics follows an effective SIAM as for the F-type exponents.

Since we use the Coulomb repulsion $U/\Gn=1 < \Uc/\Gn$ for $\Gn/D=0.01$
as in the case of F-type bath exponents, the BFAM can only access the SSC
and L phase by varying over $g$.
A larger Coulomb repulsion (not shown here) is needed to enter the LM phase
being a mirror image of the L phase.

\begin{figure}[tb]
  \includegraphics[width=\linewidth]{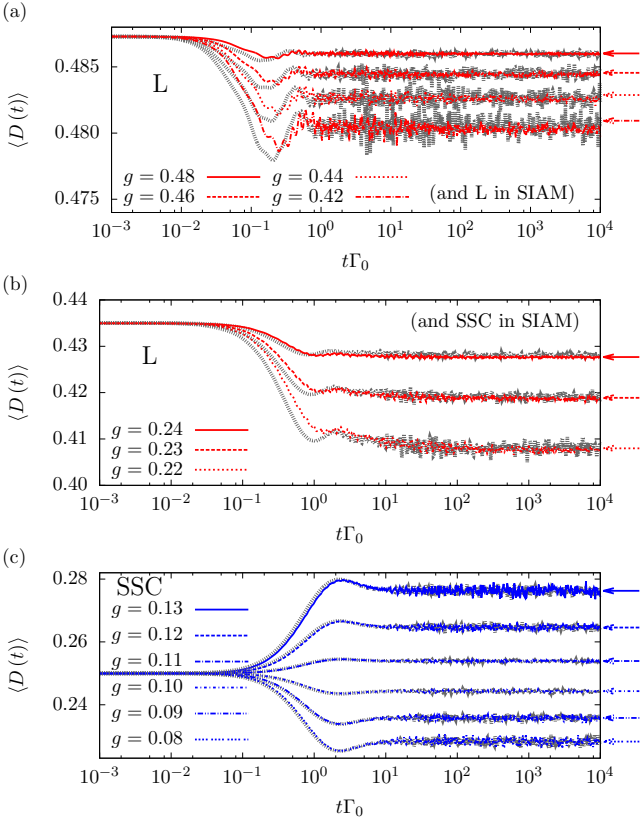}
  \caption{\label{fig:Dt-B-within}%
    (Color online)
    Real-time dynamics of the double occupancy $\Dt$ (colored curves) of
    the BFAM for bath exponents $(r,s)=(0.1,0.6)$ (B-type)
    with $U/\Gn=1$ and $\Gn/D=0.01$.
    Gray curves in the background are the corresponding double occupancy
    of the SIAM with $\Gn/D=0.01$.
    We perform quenches (a) deep within the L phase, where also the SIAM
    in the L phase, (b) in the L phase of the BFAM, while the SIAM is in
    the SSC phase, and (c) within the SSC phase of both models.
  }
\end{figure}

In the following we investigate three cases:
(a) The BFAM and the effective SIAM are in the L phase requiring a very
strong attractive Coulomb interaction in the SIAM and a large bosonic
coupling in the BFAM.
(b) For the same initial and final $\Deq$, the BFAM is characterized by
the L FP while the effective SIAM remains in SSC FP.
(c) The real-time dynamics for both models are governed by the SSC FP.

Using the parameters stated in Tab.~\ref{tab:g-U-B}, with the initial
conditions in bold, the real-time double occupancy $\Dt$ for these
three cases are plotted in Fig.~\ref{fig:Dt-B-within}.
The results for the BFAM are shown as colored curves while the
corresponding data using the effective SIAM are depicted as gray
curves in the background.

We start with quenches deep in the L phase with an initial value
$\Deq \approx 0.49$ which indicates that the local spin states
$\ket{\sigma}$ are almost completely depopulated.
This requires a strong bosonic couplings in the BFAM and a very large
attractive $\Uren$ in the SIAM.
The corresponding real-time dynamics starting and ending at the same
equilibrium expectation values $\Deq$ are shown in
Fig.~\ref{fig:Dt-B-within}(a).
Trivially, the starting and the end points are identical by construction.
However, we observe clear differences in the dynamics in both models.
The local decay minimum is much more pronounced in the SIAM while the
additional low-energy bosonic modes lead to stronger damping at all times.
Furthermore, the characteristic decay time is significantly longer in the BFAM.
Here, the effect of the bosonic bath influences directly the dynamics and
cannot be casted into a single parameter $\Uren$.
Since the bosonic coupling is large the charge fluctuation scale $\Gamma_0$
will be renormalized \cite{LangFirsov1962,Mahan81,HewsonMeyer02,
EidelsteinSchiller2012,JovchevAnders2013}.

Setting $t_0=0$ in Eq.~\eqref{eq:13} reduces the BFAM to the
Anderson-Holstein model \cite{HewsonMeyer02}.
After decoupling the fermionic bad from the impurity, the remaining
problem can be solved exactly via Lang-Firsov transformation
\cite{LangFirsov1962,Mahan81} leading to the formation of a polaron and
a shifted harmonic oscillator.
After applying the unitary Lang-Firsov transformation to the
Anderson-Holstein model, the coupling to the fermionic bath
$H_{\rm int,F}$ acquires additional factors by replacing
\begin{eqnarray}
d_\sigma^\dagger & \to & d^\dagger_\sigma 
= \bar d^\dagger_\sigma e^{ -\frac{g_0}{\omega_0} ( b_0^\dagger - b_0^{\phantom{\dagger}})}
\end{eqnarray}
where the operator $\bar{d}^\dagger_\sigma$ creates a local polaron \cite{LangFirsov1962,Mahan81,JovchevAnders2013}
This leads to a dynamic reduction of the bare hybridization strength
to a value $\Geff$ at very low temperature.
$\Geff$ depends on the RG flow, is iteration dependent
\cite{JovchevAnders2013} and only reaches a new constant value
in the low-temperature fixed point.
In the full BFAM, however, an additional energy or iteration dependency
is generated by the bosonic bath neglected in the  Anderson-Holstein model.
For small $g_0/\w_0$ the renormalization can be neglected but
for large values it can become relevant for the dynamics.

Consequently, we relate the fact that we have not been able to find a
suitable set of parameters for the SIAM with
$\expect{D_{\rm eq} (U/\Gn, D/\Gn)}=const$
and simultaneously maps the BFAM dynamics onto those of an effective SIAM
to the energy-dependent renormalization \cite{EidelsteinSchiller2012,
JovchevAnders2013} of $\Geff$ in the presence of bosonic modes.

The real-time dynamics in the BFAM depicted in Fig.~\ref{fig:Dt-B-within}(b)
is governed by the L FP while for the same initial and final values $\Deq$
the dynamic in the effective SIAM is  driven by the SSC FP:
A mapping of the BFAM on the SIAM clearly fails and the comparison between
the quenches in the BFAM and the SIAM is less fruitful since one compares
different quench types.
We note, however, that the decay time in the BFAM is also significantly
enhanced compared to the effective SIAM in this case which might be
attributed to the reduction of the charge fluctuation scale $\Geff$.

Finally, we present the dynamics for quenches within the SSC phase for both
models in Fig.~\ref{fig:Dt-B-within}(c).
We start from the locally degenerated SSC FP, \ie all four local states
are equally occupied corresponding to $\Uren = 0$ and change either the
bosonic coupling in the BFAM or quench the corresponding $\Uren$ as stated
in Tab.~\ref{tab:g-U-B} in the SIAM.
Now, the dynamics of both models coincide again.
Firstly, the bosonic coupling is significantly lower than in the regimes
depicted in Fig.~\ref{fig:Dt-B-within}(a) and (b) leading to the much weaker
renormalization of $\Geff$.
Secondly, the dynamics in the SSC phase is more dominated by the itinerant
nature of the low-energy excitations leading to the Kondo effect.

Note that also for all presented quenches with B-type bath
exponents we always observe equilibration, and the additional
bosonic environment leads to a stronger damping of the
long-time finite size oscillations compared to the SIAM.
Thermalization, however, is found  for all SSC quenches and for
L quenches as long as $|\gf - \gi|$ is not too large.

\subsection{Real-time dynamics for the quenches out of an asymmetric phase}
\label{sec:ph-symmetry-broken}

\begin{table}[b]
  \caption{\label{tab:quenches-phasym}%
    Set of final parameters for the quenches in Fig.~\ref{fig:ndt_phasym}
    in presence of particle-hole symmetry $\ed = - U/2$.
  }
  \begin{ruledtabular}
    \begin{tabular}{lccr}
                   $\Deq$  &  Phase  &      $g$      &   $U/\Gn$   \\ \colrule
  \multirow{3}{*}{0.4558}  &    L    &  $0.2 > \gc$  &     -8      \\
                           &    L    &  $0.4 > \gc$  &      6      \\
                           &   SSC   &    (SIAM)     &  $-13.144$  \\ \colrule
  \multirow{3}{*}{0.3977}  &   SSC   &  $0.2 < \gc$  &     -2      \\
                           &    L    &  $0.4 > \gc$  &     11      \\
                           &   SSC   &    (SIAM)     &   $-6.223$  \\ \colrule
  \multirow{3}{*}{0.2156}  &   SSC   &  $0.2 < \gc$  &     4.6     \\
                           &   SSC   &  $0.4 < \gc$  &     15      \\
                           &   SSC   &    (SIAM)     &   $1.1753$
    \end{tabular}
  \end{ruledtabular}
\end{table}

In the previous section we have demonstrated that the real-time dynamics
can essentially be described by an effective SIAM for F-type bath exponents
using the corresponding $\Uren$ induced by the presence of the bosonic bath.
For B-type exponents, we have observed derivations between the full dynamics
and those of the effective SIAM for large bosonic coupling constants:
The relevant time scales for the full dynamics are longer than the one of
the effective SIAM due to the energy-dependent renormalization of $\Geff$.

For ph symmetry broken models, the different asymmetric FPs have been identified
in Refs.~\cite{ChengGlossopIngersent2009,PixleyKirchnerIngersentSi2013}.
In this section, we will address the two questions:
(i) How does the system evolve from a strongly ph-symmetry broken state when
the propagation is governed by a ph-symmetric Hamiltonian $\Hf$, and
(ii) what are the differences between an effective SIAM and the full BFAM?

In order to focus on exemplifying those two questions, we restrict ourselves
to the dynamics for a constant DOS ($r=0$) and set $s=0.6$, hence we use
B-type bath exponents.
The low-temperature equilibrium phase of the BFAM with a broken ph symmetry
is characterized by the asymmetric SC FP for all $|U/\Gn| > 0$ and all
$g \geq 0$.
Similar to the spin-boson model \cite{Leggett1987,BullaTongVojta2003} no
quantum phase transition can occur in the symmetry broken state.

We prepare the system initially in the ph-symmetry broken state by applying
$|\edi/\Gn| \gg 1$ mimicking a large positive or negative gate voltage in a
quantum dot \cite{Elzerman04} which depletes or completely fills the local
orbital.
Then, we restore ph symmetry by quenching the system to $\edf = -\Uf/2$
instantaneously at $t=0$.
We leave $U=const$ at any time and only investigate the difference between
$\gi = \gf$ and artificially switching on the interaction with the bosonic
bath as well, \ie $\gi=0 \to \gf >0$.

In order to compare the effect of different values of $\gf$, below $\gf<\gc$
and above $\gf>\gc$ the critical value, and to address the question of
thermalization, we adjust the bare value of $U$ such that for each set
of parameters the same equilibrium double occupancy $\Deq$ is obtained
for all $\Hf$.
The fermionic bath is for all calculations constantly coupled with
$\Gamma(t)/D = 0.01$.

By appropriate choices of $U$, we can find two parameter pairs
$(\gf, \Uf)_{1,2}$ for a given $\Deq=const$ such that both pairs belong to
(i) the SSC phase, (ii) the L phase or (iii) are located in two different
phases such that  $g_1 < \gc$ and $g_2 > \gc$.
In addition, we determine $\Uren$ for the SIAM such that the local
thermodynamic expectation values are equal to those of the BFAM.
The FP of $\Hf$, however, is always the symmetric SC FP for the SIAM.

\begin{figure}[tb]
  \includegraphics[width=\linewidth]{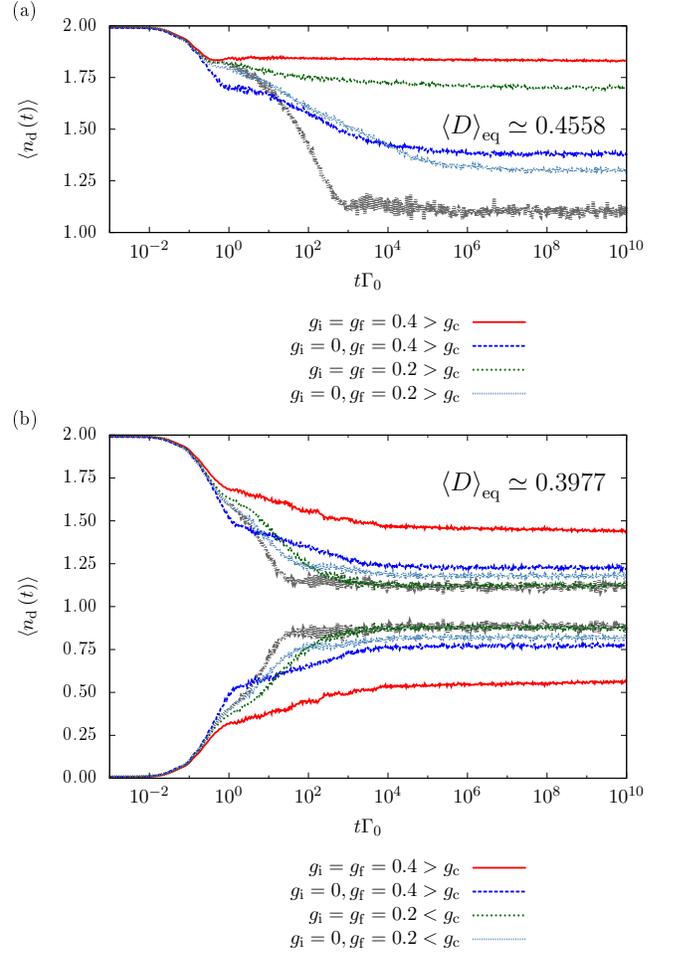}
  \caption{\label{fig:ndt_phasym}%
    (Color online)
    Dynamics of the level occupancy $\ndt$ of the BFAM with $(r,s)=(0,0.6)$
    for quenches from ph asymmetry to ph symmetry at finite temperature
    $T/D \sim 1.2 \cdot 10^{-7}$ for the equilibrium double occupancy
    (a) $\Deq \simeq 0.4558$ and (b) $\Deq \simeq 0.3977$.
    NRG parameters are $\Lambda = 6$, $\Ns = 800$, $\Nb = 8$, $\Nz = 8$
    for the BFAM and changed to $\Ns = 2000$ for the SIAM.
  }
\end{figure}

The real-time dynamics of $\ndt$ for the three representative equilibrium
double occupancies $\Deq=const$ is shown in Figs.~\ref{fig:ndt_phasym} and
\ref{fig:ndt_phasym-II}.
Independent of the model parameters, $\ndeq = 1$ due to ph symmetry.
For each plot, we have used the two coupling constants $g_1=0.2$ and $g_2=0.4$.
The corresponding values for $U$ in the three cases are summarized in
Tab.~\ref{tab:quenches-phasym}.
Additionally we supplement the dynamics of $\ndt$ obtained with the effective
SIAM for the same $\Deq$; for the $U$ values see Tab.~\ref{tab:quenches-phasym}.

We start with $\Deq \simeq 0.4558 > 1/4$ for all $\Hf$, corresponding to 
$\Uren < 0$ as stated in Tab.~\ref{tab:quenches-phasym}.
Both final values of $\gf = 0.2,0.4$ are located in the L phase of the BFAM.
For each of the final couplings we either start from an initially decoupled
bosonic bath ($\gi=0$) or the bosonic bath coupling remains unaltered
($\gi=\gf$).
The resulting real-time dynamics depicted in Fig.~\ref{fig:ndt_phasym}(a)
reveals characteristic differences in the real-time dynamics:
(i) the larger the coupling to the bosonic bath the larger the steady-state
value of $\nd^\infty$,
\begin{align}
  \nd^\infty = \lim_{T \to \infty} \frac{1}{T}\int_0^T dt \ndt
  \quad ,
\end{align}
and (ii) the bosonic system maintains its inertia for $g=const$ suppressing
the relaxation to the equilibrium expectation value even further.
In order to illustrated the difference a gray shaded curve (color online) of
the corresponding dynamics using the effective SIAM has been added for
comparison.

Note that the short-time dynamics is identical for all three cases:
The charge flow off the impurity is govern by the charge fluctuation scale
$\Gn$.
After an initial $t^2$ decay, a friction induced slowdown can be observed.
In the localized phase, an effective DOF decouples from the rest of the bath,
and therefore, its overlap with the initial state prevents the system to
approach its equilibrium value.

Our results agree with previous real-time investigations on the spin-boson model
\cite{AndersSchiller2006,AndersBullaVojta2007} where in the localized phase
the spin polarization is constant over a long stretch of time.
However, $\ndt$ remains constant even for times $t / T > 1$,
for the finite temperature $T/D \sim 1.2 \cdot 10^{-7}$ of our calculations.

This indicates that also thermal fluctuations do not influence the long-time
behavior of the level occupancy in contrary to the dynamics in the
spin-boson model.
We relate this behavior to the conservation laws of charge and spin which is
absent in the spin-boson model.
For the spin dynamics in the central spin model \cite{Gaudin1976} it is known
\cite{UhrigHackmann2014} that the existence of a non-decaying fraction
of a spin-correlation function is related to conservation laws even in
absence of integrability.
The lower bound of the long-time limit could be estimated
\cite{UhrigHackmann2014} by the application of the Mazur inequality
\cite{Mazur1969}.
In our case, however, the initial density matrix and the final states
represent a complicated many-body problem that prevents us from a straight
forward evaluation of a Mazur inequality in order to obtain an analytic
estimate for the lower boundary of $\nd^\infty$ depending on the initial
conditions.

\begin{figure}[tb]
\includegraphics[width=\linewidth]{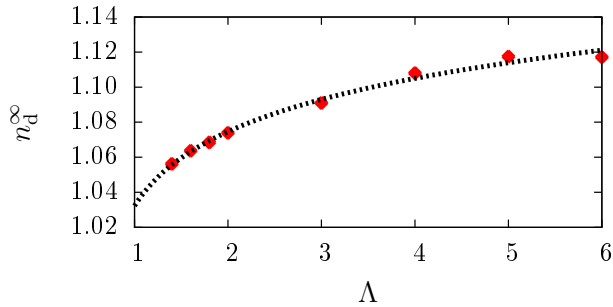}
  \caption{\label{fig:ndlong-lambda}%
    (Color online)
    $\Lambda$ dependence of $\nd^\infty$ marked as rhombuses and calculated
    for the effective SIAM with $U/\Gn = -6.223$ as used in
    Fig.~\ref{fig:ndt_phasym}(b).
    The dashed curve is an optimal fit curve extrapolating $\nd^\infty$
    in the limit $\Lambda \to 1^+$.
    NRG parameters as in Fig.~\ref{fig:ndt_phasym}.
  }
\end{figure}

Choosing $\Deq \simeq 0.3977$ enables us to find two parameter pairs
$(\gf, \Uf)$ such that for $g_1=0.2$, $\Hf$ approaches 
the SSC FP while for $g_2=0.4$, we find a L FP in equilibrium.
$\ndt$ is presented in Fig.~\ref{fig:ndt_phasym}(b) for those parameters.
For all chases, the initial decay is governed by $\Gn$, a steady state is
found at long times, and the system equilibrates.
Note that we present results on an exponentially long time scale not
accessible to other methods.

For $g_2=0.4 > \gc$, the deviations to the thermal equilibrium are large
and depend on the initial condition.
For a constant coupling to the bosonic bath, the relaxation is significantly
suppressed.
For $\gi=0$, however, the build up of correlations in the L phase needs some
time and $\nd^\infty$ is closer to the thermal value.
For $g_1=0.2 < \gc$, the steady-state value agrees perfectly with those
obtained from the effective SIAM.
Data for the two complementary cases, $\nd(t=0) \approx 0$ and
$\nd(t=0) \approx 2$ are shown to demonstrate the particle-hole
symmetry of these curves.

\begin{figure}[tb]
  \includegraphics[width=\linewidth]{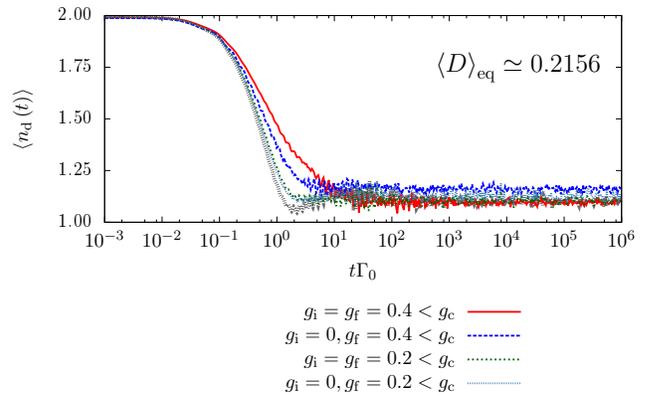}
  \caption{\label{fig:ndt_phasym-II}%
    (Color online)
    Dynamics of the level occupancy $\expect{\nd(t)}$ of the BFAM
    with $r=0$ and $s=0.6$ for quenches from ph asymmetry to ph symmetry
    at finite temperature $T/D \sim 1.2 \cdot 10^{-7}$ for $\Deq \simeq 0.2156$.
    NRG parameters as in Fig.~\ref{fig:ndt_phasym}.
  }
\end{figure}

Although, we notice a discrepancy between the steady-state value of
$\nd^\infty$ and its equilibrium value $\ndeq=1$, we can conclude that the
system is thermalizing in the SSC phase and the difference is a discretization
artifact of the TD-NRG \cite{AndersSchiller2006,
EidelsteinGuettgeSchillerAnders2012,GuettgeAndersSchiller2013}.
To prove this point, we have performed non-equilibrium calculations for
identical model parameters but different values of $\Lambda$.
The results are depicted in Fig.~\ref{fig:ndlong-lambda}.
Using a fit function, it becomes apparent that
$\lim_{\Lambda\to 1^+} \nd^\infty(\Lambda) = 1$ 
\footnote{%
  The small difference to $\nd^\infty = 1$ is related to the finite number
  of state $\Ns$ used in each calculation.
}.

The results for the last case $g_1,g_2 < \gc$ with $\Deq \simeq 0.2156$
are shown in Fig.~\ref{fig:ndt_phasym-II} for $\Lambda = 6$.
Within the numerical accuracy thermalization is found for all different
parameters.
Again, the initial decay is equal for all bosonic coupling constants and
only governed by $\Gn$ as before.
The steady-state value is found up to $t\Gn = 10^{10}$ although only
data up to $t\Gn=10^{6}$ are depicted in the figure.
After a fast initial decay, we note a slowdown of the dynamics in the
presence of a large coupling to the bosonic bath.

\section{Conclusion}
\label{sec:conclusion}

In this paper we have investigated the influence of a bosonic bath onto the
local non-equilibrium dynamics in the BFAM after a sudden change of parameters.
The additional bosonic bath induces an additional attractive Coulomb
interaction.
We have defined a renormalized $\Uren$ by equating the local double occupancy
in the BFAM in the presence of a finite bath coupling and the bare value
of $U$ within an effective SIAM to estimate the $\Uren$.

We have shown that the real-time dynamics for F-type bath exponents can be
fully understood within the effective SIAM defined by $\Uren$ and the
identical fermionic bath coupling function when ph symmetry is maintained
at all times.
This includes quenches across the QCP into the localized phase, which also
is found for a suitable $\Uren < 0$.
There exist a one-to-one correspondence of the QPT between the spin Kondo
effect and the local moment phase to the charge Kondo effect and the
localized phase.
The spin Kondo regime and the charge Kondo regime are adiabatically connected,
and there exists one unique SSC FP with identical spin and charge moments,
$\mueff$ and $\Qeff$ respectively.
Within the numerical accuracy, we find perfect agreement between the dynamics
in BFAM and those within the SIAM.

For B-type bath exponents a different picture emerges.
Although the major effect of the bosonic bath is included into the
renormalization $U \to \Uren$, the non-equilibrium dynamics of the full
BFAM differs significantly to those of the effective SIAM.
While for the quenches within the SSC phase the dynamics in both modes coincide,
the mapping of the dynamics of the BFAM onto an effective SIAM fails for
quenches within the L phase:
Either in the effective SIAM the quench is of a different type, while the
BFAM is already driven by the bosonic bath, or the bosonic coupling is very
strong so that  the characteristic time scale of the dynamics is increased
in the BFAM compared to the dynamics in the SIAM.

Although all quenches equilibrate for long times to a steady-state value,
thermalization is only found for quenches into the SSC phase or
very close to the QCP where the characteristic low energy scale vanishes.
With increasing distance to the QCP, the low-energy LM or L FP is approached
faster in the NRG iterations, indicating a smaller spatial extension
\cite{Wilson75,Barzykin1996,LechtenbergAnders2014,KleineMusshoffAnders2014} 
of the decoupled effective moment.
Then, the difference between the steady-state value and the thermal equilibrium
increases, since the local operators have a greater overlap with the decoupled
local moments the larger the characteristic crossover energy scale to the
low-temperature FP becomes.

Furthermore, we have presented data for quenches from a ph-symmetry broken
initial phase to a ph-symmetric state.
There, we have restricted ourselves to $r=0$ and have focused only on the
effect of an additional bosonic environment.
In this case the LM phase is absent and the BFAM only shows a QPT between
the SSC and the L phase.

The  level occupancy $\ndt$ always approaches a steady-state value for all
investigated parameter sets.
Most strikingly, however, $\nd^{\infty}$ deviates strongly from its
thermodynamic equilibrium value, calculated with the final Hamiltonian,
for quenches into the L phase ($g > \gc$).
This is related to a decoupling of an effective charge DOF from the bath
continuum, preventing the local system to relax to the thermodynamic
equilibrium.
The magnitude of the deviation between the steady-state value and the
equilibrium value depends on the initial condition.
For an initially decoupled bosonic bath ($\gi=0$) the development of the
localized charge DOF takes longer than the charge flow, and consequently,
the difference is smaller compared to the case of $\gi=\gf$.

For quenches within the SSC phase, $g < \gc$, the dynamics of $\ndt$ in the
BFAM and the effective SIAM are similar, although the presence of the
bosonic coupling yields a slowdown of the relaxation process.
Nevertheless, $\ndt$ thermalizes to its equilibrium value for long times
in all cases $g < \gc$, which we have shown by an extrapolation of
$\nd^\infty(\Lambda)$ for $\Lambda \to 1^+$.

\begin{acknowledgments}
We would like to thank F.\ G\"{u}ttge for his valuable contributions
to the Bose-Fermi NRG code.
We acknowledge financial support by the Deutsche Forschungsgemeinschaft
through AN 275/7-1.
\end{acknowledgments}

%

\end{document}